%% file: florit-smith.tex
\documentclass{conm-p-l}

\usepackage{amssymb}

\usepackage{graphicx}

\usepackage{amsmath,amsfonts}
\usepackage{amsthm}
\usepackage{mathtools}
\usepackage{multirow}
\usepackage{tikz}
\usetikzlibrary{calc}
\usetikzlibrary{shapes.geometric}
\usetikzlibrary{arrows}
\usetikzlibrary{automata}
\usetikzlibrary{positioning}
\usepackage{url}
\usepackage{hyperref}
\usepackage{doi}
\usepackage{xcolor}
\usepackage{booktabs}
\usepackage{xspace}
\usepackage[all,cmtip]{xy}
\usepackage[ruled,vlined,linesnumbered]{algorithm2e}
\usepackage{fancyvrb}
\DontPrintSemicolon
\usepackage{subfig}
\usepackage{longtable}
\usepackage{xargs}

\newtheorem{theorem}{Theorem}[section]
\newtheorem{lemma}[theorem]{Lemma}
\newtheorem{proposition}[theorem]{Proposition}
\newtheorem{corollary}[theorem]{Corollary}

\theoremstyle{definition}
\newtheorem{definition}[theorem]{Definition}
\newtheorem{example}[theorem]{Example}

\newtheorem{conjecture}[theorem]{Conjecture}

\theoremstyle{remark}
\newtheorem{remark}[theorem]{Remark}

\numberwithin{equation}{section}

\newcommand{\PP}{\ensuremath{\mathbb{P}}}
\newcommand{\FF}{\ensuremath{\mathbb{F}}}
\newcommand{\FFbar}{\ensuremath{\overline{\mathbb{F}}}}

\newcommand{\ZZ}{\ensuremath{\mathbb{Z}}}
\newcommand{\RR}{\ensuremath{\mathbb{R}}}
\newcommand{\XC}{\ensuremath{\mathcal{C}}}
\newcommand{\EC}{\ensuremath{\mathcal{E}}}
\newcommand{\AV}{\ensuremath{\mathcal{A}}}

\newcommand{\field}{\ensuremath{\Bbbk}\xspace}

\newcommand{\Jac}[1]{\ensuremath{\mathcal{J}({#1})}}
\newcommand{\Aut}{\ensuremath{\mathrm{Aut}}}
\newcommand{\RAut}{\ensuremath{\mathrm{RA}}}

\newcommand{\dualof}[1]{\ensuremath{#1^\dagger}}

\newcommand{\classof}[1]{\ensuremath{\big[{#1}\big]}}

\newcommand{\subgrp}[1]{\ensuremath{\langle{#1}\rangle}}

\newcommand{\Prob}[1]{\ensuremath{\operatorname{Pr}\left[#1\right]}}

\newcommand{\softO}{\ensuremath{\widetilde{O}}}

\newcommandx{\IG}[3][1=g, 2=\ell, 3=p]{\ensuremath{\Gamma_{#1}(#2;#3)}\xspace}
\newcommandx{\SSIG}[3][1=g, 2=\ell, 3=p]{\ensuremath{\Gamma^{SS}_{#1}(#2;#3)}\xspace}
\newcommand{\RichelotIG}{\ensuremath{\Gamma^{SS}_{2}(2;p)}\xspace}

\newcommand{\TypeA}{\textsf{Type-A}\xspace}
\newcommand{\TypeI}{\textsf{Type-I}\xspace}

\newcommand{\TypeII}{\textsf{Type-II}\xspace}
\newcommand{\TypeIII}{\textsf{Type-III}\xspace}

\newcommand{\TypeIV}{\textsf{Type-IV}\xspace}

\newcommand{\TypeV}{\textsf{Type-V}\xspace}
\newcommand{\TypeVI}{\textsf{Type-VI}\xspace}

\newcommand{\TypeExE}{{\textsf{Type-$\Pi$}}\xspace}
\newcommand{\TypeEsquared}{{\textsf{Type-$\Sigma$}}\xspace}
\newcommand{\TypeExEzero}{{\textsf{Type-$\Pi_0$}}\xspace}
\newcommand{\TypeEzerosquared}{{\textsf{Type-$\Sigma_0$}}\xspace}
\newcommand{\TypeExEtwelvecubed}{{\textsf{Type-$\Pi_{12^3}$}}\xspace}
\newcommand{\TypeEtwelvecubedsquared}{{\textsf{Type-$\Sigma_{12^3}$}}\xspace}
\newcommand{\TypeEzeroxEtwelvecubed}{{\textsf{Type-$\Pi_{0,12^3}$}}\xspace}

\begin{document}

\title[Automorphisms and Superspecial Richelot Isogeny
Graphs]{Automorphisms and isogeny graphs of abelian varieties,\\ with
applications to the superspecial Richelot isogeny graph}

\author{Enric Florit}
\address{IMUB - Universitat de Barcelona, Gran Via de les Corts Catalanes 585, 08007 Barcelona, Spain}
\email{efz1005@gmail.com}

\author{Benjamin Smith}
\address{Inria and Laboratoire d'Informatique (LIX), CNRS, École
polytechnique, Institut Polytechnique de Paris, 91120 Palaiseau, France}
\email{smith@lix.polytechnique.fr}

\thanks{The second author was supported in part by l'Agence nationale de
la recherche (ANR) program CIAO ANR-19-CE48-0008.}

\subjclass[2020]{Primary 14K02; Secondary 14G50, 14Q05, 11T99, 05C81}

\date{}

\keywords{Superspecial abelian varieties, isogeny graphs, isogeny-based cryptography}

\begin{abstract}
    We investigate special structures due to automorphisms
    in isogeny graphs of principally
    polarized abelian varieties,
    and abelian surfaces in particular.
    We give theoretical and experimental results
    on the spectral and statistical properties
    of \((2,2)\)-isogeny graphs of superspecial abelian surfaces,
    including stationary distributions for random walks,
    bounds on eigenvalues and diameters,
    and a proof of the connectivity of the Jacobian subgraph
    of the \((2,2)\)-isogeny graph.
    Our results improve our understanding of the performance and
    security of some recently-proposed cryptosystems,
    and are also a concrete step towards a better understanding
    of general superspecial isogeny graphs in arbitrary dimension.
\end{abstract}

\maketitle

\input introduction

\input automorphisms

\input random-walks

\input ppas

\input random-richelot

\input connectivity

\input example-p-47


\bibliographystyle{amsplain}
\bibliography{references}

\appendix

\input data

\input explicit

\end{document}

%% file: introduction.tex
\section{
    Introduction
}

When studying the internal structure of isogeny classes of abelian varieties
from an algorithmic point of view,
we work with \emph{isogeny graphs}:
the vertices are isomorphism classes of abelian varieties,
and the edges are isomorphism classes of isogenies,
often of some fixed degree.
For elliptic curves,
these graphs have already had a wealth of applications.
Mestre~\cite{1986/Mestre}
used his \emph{méthode des graphes} 
to compute a basis of the space \(S_2(N)\) 
of modular forms of weight 2, level \(N\), and trivial character.
Kohel~\cite{1996/Kohel} used isogeny graphs
to compute endomorphism rings of elliptic curves over finite fields,
and Fouquet and Morain turned this around to improve
point-counting algorithms for elliptic curves~\cite{2002/Fouquet--Morain}.
Bröker, Lauter, and Sutherland~\cite{2012/Broker--Lauter--Sutherland}
developed an algorithm for computing modular polynomials
using isogeny graph structures;
Sutherland~\cite{2012/Sutherland}
has used the difference between the structures of ordinary and
supersingular isogeny graphs to give a remarkable and efficient
deterministic supersingularity test for elliptic curves.

More recently, isogeny graphs have become a setting for post-quantum
cryptographic algorithms, especially in the supersingular case.
Charles, Goren, and Lauter proposed a cryptographic hash function
with provable security properties based on combinatorial properties
of the supersingular elliptic \(2\)-isogeny graph~\cite{2009/CGL}.
Rostovtsev and Stolbunov proposed a key exchange scheme 
based on ordinary isogeny
graphs~\cite{2006/Rostovtsev--Stolbunov,2010/Stolbunov};
this was vastly accelerated by Castryck, Lange, Martindale, Panny, and Renes 
by transposing it to a subgraph of the supersingular isogeny graph, 
where it is known as CSIDH~\cite{2018/CLMPR}.
Jao and De Feo's SIDH key exchange
algorithm~\cite{2011/Jao--De-Feo,2014/DJP},
the basis of SIKE~\cite{2017/SIKE}
(a third-round alternate candidate 
in the NIST post-quantum cryptography standardization process),
is based on the difficulty of finding paths in the elliptic supersingular
\(2\)- and \(3\)-isogeny graphs.
These applications all depend, both in their constructions and in
their security arguments, on a precise understanding 
of the combinatorial properties of supersingular isogeny graphs.

It is natural to try to extend these applications
to the setting of isogeny graphs of 
higher-dimensional principally polarized abelian varieties (PPAVs).
First steps in this direction have been made by
Charles, Goren, and Lauter~\cite{2009/CGLgenusg},
Takashima~\cite{2018/Takashima},
Flynn and Ti~\cite{2019/Flynn--Ti},
and
Castryck, Decru, and Smith~\cite{2020/Castryck--Decru--Smith}.
Costello and Smith have proposed an attack on cryptosystems 
based on the difficulty of computing isogenies 
between higher-dimensional superspecial abelian
varieties~\cite{2020/Costello--Smith}.

But so far, the efficiency and security of these algorithms
is conjectural---even speculative---because 
of a lack of information on combinatorial properties of
supersingular isogeny graphs in higher dimension,
such as their connectedness, their diameter, and their expansion constants.
For example, the hash functions
typically depend on the rapid convergence of random walks to the uniform
distribution on the isogeny graph;
but while this is well-known for the elliptic case,
it is not yet well-understood even in \(g = 2\).
Indeed, even the connectedness of the superspecial graph for \(g = 2\)
has only recently been proven 
by Jordan and Zaytman~\cite{2020/Jordan--Zaytman}.

Our ultimate aim is a deeper 
understanding of the combinatorial and
spectral properties of the superspecial graph,
such as its diameter and the limit distribution 
of random walks.
In this article we give some theoretical results on general superspecial graphs,
and experimental results focused on the 
\emph{Richelot isogeny graph}:
that is, the graph formed by \((2,2)\)-isogenies
of \(2\)-dimensional PPAVs.
Richelot isogeny graphs are the most amenable to explicit computation
(apart from elliptic graphs),
and already exhibit a particularly rich structure.

After recalling basic results in~\S\ref{sec:basics},
we explore the impact of automorphisms of \(g\)-dimensional PPAVs 
on edge weights in the \((\ell,\ldots,\ell)\)-isogeny graph
for general \(g\) and \(\ell\)
in~\S\ref{sec:automorphisms}.
Automorphisms are a complicating factor 
that can almost be ignored in elliptic isogeny graphs,
since only two vertices (corresponding to \(j\)-invariants 0 and 1728)
have automorphisms other than \(\pm1\).
In higher dimensions, however,
extra automorphisms are much more than an isolated corner-case:
every general product PPAV \(\mathcal{A}\times\mathcal{B}\)
has an involution \([1]_{\mathcal{A}}\times[-1]_\mathcal{B}\)
which may induce nontrivial weights in the isogeny graph,
and entire families of simple PPAVs can come equipped with
extra automorphisms,
as we will see in~\S\ref{sec:ppas} for dimension \(g = 2\).
The \emph{ratio principle} proven in Lemma~\ref{lemma:key},
which relates automorphism groups of \((\ell,\ldots,\ell)\)-isogenous PPAVs
with the weights of the directed edges between them in the isogeny
graph, is an essential tool for our later investigations.

We consider the spectral and statistical properties of isogeny graphs,
still in the most general setting, in~\S\ref{sec:random-walks}.
Here we prove results which, combined with an understanding of the
automorphism groups of vertices, allow us to state general theoretical
bounds on eigenvalues, and compute stationary distributions
for random walks in the superspecial isogeny graph---and also in
interesting subgraphs of the superspecial graph,
such as the Jacobian subgraph.

We then narrow our focus to the Richelot isogeny graph:
that is, the case \(g = 2\) and \(\ell = 2\).
We recall Bolza's classification of automorphism groups
of genus-2 Jacobians in~\S\ref{sec:ppas},
and apply it in the context of Richelot isogeny graphs
(extending the results of Katsura and
Takashima~\cite{2020/Katsura--Takashima}).
In~\S\ref{sec:random-Richelot}
we specialize our general results to \(g = 2\) and \(\ell = 2\),
and give experimental data 
for diameters and second eigenvalues
of superspecial Richelot isogeny graphs
(and Jacobian subgraphs)
for \(17 \le p \le 601\).
This allows us to prove that the Jacobian subgraph 
of the Richelot isogeny graph
is connected and aperiodic,
and to bound its diameter
relative to the diameter of the entire superspecial graph
in~\S\ref{sec:connectivity}.

Our results have consequences for the security and efficiency arguments
of the cryptographic algorithms described
in~\cite{2018/Takashima},
\cite{2019/Flynn--Ti},
\cite{2020/Castryck--Decru--Smith},
and~\cite{2020/Costello--Smith}.
For example,
we can estimate the frequency with which elliptic products are
encountered during random walks in the superspecial graph,
which is essential for understanding the true efficiency
of the attack in~\cite{2020/Costello--Smith};
and we can understand the stationary distribution for random walks
restricted to the Jacobian subgraph
(which were used in~\cite{2020/Castryck--Decru--Smith}).
These cryptographic implications are further discussed
in~\S\ref{sec:random-Richelot}.
Our results also offer a concrete step towards a better understanding
of the situation for general superspecial isogeny graphs---that is,
in arbitrary dimension \(g\), 
and with \((\ell,\ldots,\ell)\)-isogenies
for arbitrary primes~\(\ell\).

\section{
    Isogeny graphs
}
\label{sec:basics}

\begin{definition}
    Let \(\AV/\field\) be a principally polarized abelian variety (PPAV) and
    \(\ell\) a prime,
    not equal to the characteristic of \(\field\).
    A subgroup of \(\AV[\ell]\) is \textbf{Lagrangian}
    if it is maximally isotropic with respect to the \(\ell\)-Weil pairing.
    An \textbf{\((\ell,\ldots,\ell)\)-isogeny} is an isogeny \(\AV \to \AV'\) of
    PPAVs
    whose kernel is a Lagrangian subgroup of \(\AV[\ell]\).
\end{definition}

If \(\AV\) is a \(g\)-dimensional PPAV,
then every Lagrangian subgroup of \(\AV[\ell]\)
is necessarily isomorphic to \((\ZZ/\ell\ZZ)^g\),
though the converse does not hold.
Since its kernel is Lagrangian,
an \((\ell,\ldots,\ell)\)-isogeny \(\phi: \AV \to \AV'\)
respects the principal polarizations:
if \(\lambda\) and \(\lambda'\)
are the principal polarizations on \(\AV\) and \(\AV'\),
respectively,
then the pullback \(\phi^*(\lambda')\) 
is equal to \(\ell\lambda\).

Given another $g$-dimensional PPAV $\AV'$, we say two Lagrangian subgroups $K$ of $\AV[\ell]$ and $K'$ of $\AV'[\ell]$ yield \textbf{isomorphic isogenies} $\phi$ and $\phi'$, if there are isomorphisms $\alpha \colon \AV \to \AV'$ and $\beta\colon \AV/K \to \AV/K'$ respecting the principal polarizations, such that the following diagram commutes:
\[
\xymatrix{
\AV \ar[r]^{\alpha} \ar[d]_{\phi} & \AV' \ar[d]^{\phi'}\\
\AV/K \ar[r]_{\beta} & \AV'/K'
}
\]
In this case, the dual isogenies $\dualof{\phi}$ and $\dualof{\phi'}$ are also isomorphic.

\begin{definition}
    Fix a positive integer $g$ and a prime $p$. The \textbf{\((\ell,\dots,\ell)\)-isogeny graph}, denoted \(\IG\), is the directed weighted multigraph defined as follows.
    \begin{itemize}
        \item 
            The \textbf{vertices} are isomorphism classes of
            PPAVs defined over \(\bar\FF_p\).
            If \(\AV\) is a PPAV,
            then \(\classof{\AV}\) denotes the corresponding vertex.
        \item
            The \textbf{edges} are isomorphism classes of
            \((\ell, \dots, \ell)\)-isogenies,
            \textbf{weighted} by the number of distinct kernels yielding
            isogenies in the class.
            The weight of an edge \(\classof{\phi}\)
            is denoted by \(w(\classof{\phi})\).
    \end{itemize}
\end{definition}

If \(\classof{\phi}: \classof{\AV}\to \classof{\AV'}\) is an edge,
then $w(\classof{\phi}) = n$
if and only if 
there are \(n\) Lagrangian subgroups \(K \subset \AV[\ell]\) 
such that \(\AV' \cong \AV/K\)
(this definition is independent of the choice of
representative isogeny \(\phi\)).
Equivalently,
if there is an \((\ell,\ldots,\ell)\)-isogeny \(\phi: \AV \to \AV'\),
then \(w(\classof{\phi})\)
is equal to the size of the orbit of \(\ker\phi\)
under the action of \(\Aut(\AV)\) on the set of Lagrangian
subgroups of \(\AV[\ell]\).

The isogeny graph breaks up into components;
there are at least as many connected components
as there are isogeny classes over \(\field\).
We are particularly interested in the superspecial isogeny class.

\begin{definition}
    A PPAV \(\AV/\FFbar_p\) of dimension \(g\)
    is \textbf{superspecial}
    if its Hasse--Witt matrix vanishes identically.
    Equivalently,
    \(\AV\) is superspecial
    if it is isomorphic \emph{as an unpolarized abelian variety}
    to a product of supersingular elliptic curves.
\end{definition}

For general facts and background on superspecial and supersingular
abelian varieties, 
we refer to Li and Oort~\cite{1998/Li--Oort},
and Brock's thesis~\cite{1993/Brock} (especially for \(g \le 3\)).

\begin{definition}
    The \((\ell,\ldots,\ell)\)-isogeny graph
    of \(g\)-dimensional superspecial PPAVs
    over \(\FFbar_p\)
    is denoted by \(\SSIG\).
    We often refer to \(\SSIG\)
    as the \emph{superspecial graph},
    with \(g\), \(\ell\), and \(p\) implicit.
\end{definition}

The graph \SSIG is regular (every vertex has the same weighted
out-degree), and Jordan and Zaytman recently proved that \SSIG is
connected
(see~\cite{2020/Jordan--Zaytman}; 
though this result was already implicit,
in a different language,
in~\cite[Lemma 7.9]{2001/Oort}).
If an elliptic curve is supersingular,
then it is isomorphic to a curve defined over \(\FF_{p^2}\).
Similarly,
if \(\AV/\FFbar_p\)
is superspecial,
then \(\AV\) is isomorphic to a PPAV defined over~\(\FF_{p^2}\),
so in our experiments involving superspecial graphs,
we work over \(\FF_{p^2}\) for various \(p\).

%% file: automorphisms.tex
\section{
    Isogenies and automorphisms
}
\label{sec:automorphisms}

Isogeny graphs are weighted directed graphs,
and before going any further,
we should pause to understand the weights.
The weights of the edges 
are closely related to the automorphism groups of the vertices that they
connect, as we shall see.

Let \(\AV\) be a PPAV,
let \(K\) be a Lagrangian subgroup of \(\AV[\ell]\) for some \(\ell\),
and let \(\alpha\) be an automorphism of \(\AV\).
We write \(K_\alpha\) for \(\alpha(K)\).

If \(K_\alpha = K\),
then \(\alpha\) induces an automorphism of \(\AV/K\).
Going further,
if \(S\) is the stabiliser of \(K\) in \(\Aut(\AV)\),
then \(S\) induces an isomorphic subgroup \(S'\) of \(\Aut(\AV/K)\).

Now suppose that \(K_\alpha \not= K\).
If \(\phi: \AV \to \AV/K\)
and \(\phi_\alpha: \AV \to \AV/K_\alpha\)
are the quotient isogenies,
then \(\alpha\) induces an isomorphism
\(\alpha_*: \AV/K \to \AV/K_\alpha\)
such that 
\(\alpha_*\circ\phi = \phi'\circ\alpha\).
(Note that \(\phi\) and \(\phi_\alpha\) are only defined up to
isomorphism, but if we fix a choice of \(\phi\) and \(\phi_\alpha\),
then \(\alpha_*\) is unique.)
%
%
%
Let \(\phi_\alpha = \alpha_*^{-1}\circ\phi'\).
The isogenies
\(\phi\) and \(\phi_\alpha\)
have identical domains and codomains, but distinct kernels;
thus, they both represent the same edge in the isogeny graph,
and \(w(\classof{\phi}) > 1\).
Going further,
if \(O_K\) is the orbit of \(K\) under \(\Aut(\AV)\),
then there are \(\#O_K\) distinct kernels
of isogenies representing~\(\classof{\phi}\):
that is, \(w(\classof{\phi}) = \#O_K\).

Looking at the dual isogenies,
we see that \(\alpha^{-1}\circ\dualof{(\phi_{\alpha})}\circ\phi = [\ell]_{\AV}\),
so \(\dualof{\phi}\) and \(\dualof{\phi_{\alpha}}\) 
have the same kernel.
Hence, while automorphisms of \(\AV\) may lead to increased weight on
the edge \(\classof{\phi}\),
they have no effect on the weight of the dual edge
\(\classof{\dualof{\phi}}\).

Every PPAV has a nontrivial involution \([-1]\),
but \([-1]\) fixes every kernel and commutes with
every isogeny.
It therefore has no impact on edges or weights in the isogeny
graph, so can simplify our analysis by quotienting it away.
Indeed, since \(\subgrp{[-1]}\) is contained in the centre of \(\Aut(\AV)\),
the quotient \(\Aut(\AV)/\subgrp{[-1]}\) acts on the set of Lagrangian subgroups of
\(\AV[\ell]\).  This is crucial in what follows.

\begin{definition}
    If \(\AV\) is a PPAV,
    then its \textbf{reduced automorphism group}\footnote{
        Reduced automorphism groups are usually defined for hyperelliptic curves,
        not abelian varieties,
        but if \(\AV = \Jac{\XC}\)
        is the Jacobian of a hyperelliptic curve
        and \(\iota\) is the hyperelliptic involution,
        then \(\RAut(\Jac{\XC})\) is
        canonically isomorphic to \(\RAut(\XC) = \Aut(\XC)/\subgrp{\iota}\);
        so our definition is consistent for hyperelliptic Jacobians.
    }
    is 
    \[
        \RAut(\AV) := \Aut(\AV)/\subgrp{[-1]}
        \,.
    \]
\end{definition}

\begin{lemma}
    \label{lemma:key}
    Let \(\phi: \AV \to \AV'\) 
    be an \((\ell,\ldots,\ell)\)-isogeny,
    and let \(S\) be the stabiliser of \(\ker(\phi)\) in \(\RAut(\AV)\).
    
    \begin{enumerate}
        \item
            The isogeny \(\phi\) induces a subgroup \(S'\) of
            \(\RAut(\AV')\) isomorphic to \(S\),
            and \(S'\) is the stabiliser of \(\ker{\dualof{\phi}}\) in \(\RAut(\AV')\).
        \item
            If \(s := \#S\) (so \(s = \#S'\)),
            then in the \((\ell,\ldots,\ell)\)-isogeny graph we have
            \[
                w(\classof{\phi}) =
                \#\RAut(\AV)/s
                \qquad
                \text{and}
                \qquad
                w(\classof{\dualof{\phi}}) =
                \#\RAut(\AV')/s
                \,.
            \]
            In particular,
            \begin{equation}
                \label{eq:ratio-principle}
                \#\RAut(\AV)\cdot w(\classof{\dualof{\phi}})
                =
                \#\RAut(\AV')\cdot w(\classof{\phi})
                \,.
            \end{equation}
    \end{enumerate}
\end{lemma}
\begin{proof}
    Let \(K := \ker(\phi)\) be the kernel of \(\phi\).
    As discussed above, 
    each \(\alpha\) in \(\Aut(\AV)\) 
    induces an isomorphism \(\alpha_*\colon \AV' \to \AV/\alpha(K)\),
    and if \(\alpha\) stabilises \(K\), then \(\alpha_*\) is an automorphism of \(\AV'\). 
    As \(\alpha\) stabilises \(\AV[\ell]\),
    this gives an inclusion of \(S\) 
    into the stabiliser of \(\ker\dualof{\phi}\).
    The reverse inclusion comes from the symmetric
    argument on the dual. 
    The second statement follows from the orbit-stabiliser theorem. Note we only need to consider the action by reduced automorphisms, as \([-1]\) acts trivially on all subgroups of \(\AV\).
\end{proof}

To understand the isogeny graph, then,
we need to understand the reduced automorphism groups
of its vertices.
A generic PPAV \(\AV\) has \(\Aut(\AV) = \subgrp{[-1]}\),
so \(\RAut(\AV) = 1\).
The simplest examples of nontrivial reduced automorphism groups
are the elliptic curves with \(j\)-invariants~\(0\) and~\(1728\).
Moving into higher dimensions,
nontrivial reduced automorphism groups are much more common:
for example, if \(\AV = \EC\times\EC'\)
is a product of elliptic curves,
then \([1]_{\EC}\times[-1]_{\EC'}\)
is a nontrivial involution in \(\RAut(\EC\times\EC')\).
We will see many more examples of nontrivial reduced automorphism groups
below. 

\begin{figure}[ht]
    \centering
    \begin{tikzpicture}[
            >={angle 60},
            thick,
            vertex/.style = {circle, draw, fill=white, inner sep=0.5mm, minimum size=9mm},
            fontscale/.style = {font=\relsize{#1}}
        ]
        
    \node[vertex] at (1.5,1.2) (C1) {$\AV_2$};
    \node[vertex] at (-1.5,1.2) (C2) {$\AV_1$};
    \node[vertex] at (-3,-1.2) (E2) {$\EC_{1728}^2$};
    \node[vertex] at (0,-1.2) (E3) {$\EC_{0}^2$};
    \node[vertex, fontscale=0.2] at (3,-1.2) (E2xE3) {$\Pi$};
    
    \node[] at (-2.2,1.6) {$3$};
    
    \node[] at (-0.9,1.7) {$2$};
    \node[] at (0.9,1.7) {$1$};
    
    \node[] at (-0.9,0.75) {$6$};
    \node[] at (0.9,0.75) {$3$};
    
    \node[] at (2.2,1.6) {$6$};
    
    \node[] at (-3.8,-0.8) {$3$};
    
    \node[] at (3.7,-0.8) {$6$};
    
    \node[] at (0.7,-0.8) {$3$};
    
    \node[] at (-2.2,-0.9) {$4$};
    \node[] at (-0.8,-0.9) {$9$};
    
    \node[] at (-2.4,-1.8) {$4$};
    \node[] at (2.4,-1.8) {$3$};
    
    \node[] at (2.9,-0.5) {$6$};
    \node[] at (2.2,0.6) {$3$};
    
    \node[] at (-2.9,-0.5) {$4$};
    \node[] at (-2.2,0.6) {$3$};
    
    \node[] at (-0.1,-0.5) {$3$};
    \node[] at (-0.6,0.4) {$1$};
    
    \draw[-] 
        (C1) edge [bend left=20] (C2)
        (C1) edge [bend right=20] (C2)
        (C1) edge (E2xE3)
        (C2) edge (E2)
        (C2) edge (E3)
        (E2) edge (E3)
        (E2) edge [bend right] (E2xE3)
        (C1) edge [out=345, in=15, looseness=8] (C1)
        (C2) edge [out=165, in=195, looseness=8] (C2)
        (E2xE3) edge [out=15, in=-15, looseness=8] (E2xE3)
        (E3) edge [out=15, in=-15, looseness=8] (E3)
        (E2) edge [out=165, in=195, looseness=8] (E2);
    \end{tikzpicture}
    \caption{The graph $\SSIG[2][2][11]$, with isogeny weights.}
    \label{fig:p-11}
\end{figure}
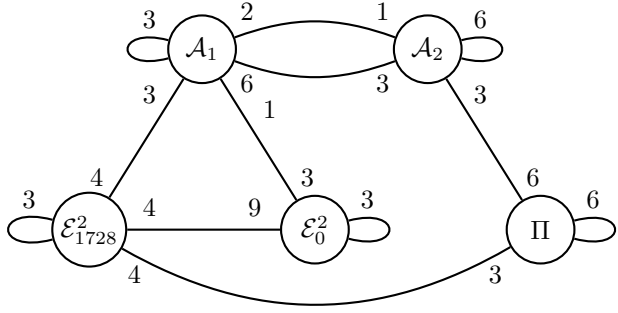

\begin{example}
    Consider the graph $\SSIG[2][2][11]$, shown in
    Figure~\ref{fig:p-11}. It has five vertices:
    \begin{itemize}
        \item 
            \(\classof{\AV_1} = \classof{\Jac{\XC_1}}\), for 
            \(\XC_1 \colon y^2 = x^6 - 1\), with 
            \(\RAut(\AV_1) = D_{2\times 6}\).
        \item \(\classof{\AV_2} = \classof{\Jac{\XC_2}}\), for 
            \(\XC_2 \colon y^2 = (x^3 - 1)(x^3 - 3)\), with 
            \(\RAut(\AV_2) = S_3\).
        \item 
            \(\classof{\EC^2_{1728}}\), where $\EC_{1728} \colon y^2 = x^3 - x$, and $\#\RAut(\EC^2_{1728}) = 16$.
        \item 
            \(\classof{\EC^2_{0}}\), where $\EC_{0} \colon y^2 = x^3 - 1$, and $\#\RAut(\EC^2_{0}) = 36$.
        \item 
            \(\classof{\Pi} = \classof{\EC_0\times \EC_{1728}}\), with $\#\RAut(\Pi) = 12$.
    \end{itemize}
    The weights indicated in the figure indeed satisfy
    Equation~\eqref{eq:ratio-principle}. For instance, there is a unique
    \((2,2)\)-isogeny  $\phi\colon \EC^2_{1728} \to \EC_{0}^2$ (up to isomorphism), and
    \[
        \frac{w(\classof{\phi})}{w(\classof{\dualof{\phi}})} 
        = \frac{4}{9} = \frac{16}{36} 
        = \frac{\#\RAut(\EC^2_{1728})}{\#\RAut(\EC^2_{0})}.
    \]
\end{example}

%% file: random-walks.tex
\section{Random walks}
\label{sec:random-walks}

Let $G = (V, E, w)$ be a directed weighted multigraph with finite vertex
set $V$.  The weight of an edge $e$ is denoted by $w(e) > 0$.
Given subsets $S, T \subset V$,
we denote the multiset of edges from $S$ to $T$ by $E(S,T)$, 
omitting the curly braces when $S$ or $T$ is a singleton $\{u\}$.
For each pair of vertices $u,v\in V$ we write $w_{uv} = \sum_{e\in
E(u,v)} w(e)$, and for each vertex $u\in V$ we have $\deg u = \sum_{e\in
E(u, V)} w(e)$. The set of neighbors of a vertex $u\in V$ (that is, the
set of vertices \(v\) such that \(E(u,v) \neq \emptyset\))
is denoted~$N(u)$.

We define a \textbf{random walk} on $G$ with starting vertex $v_0\in V$
in the usual way:
for each natural $t \geq 0$ and pair of vertices $u,v\in V$, we have 
\[P(v_{t+1} = v \mid v_t = u) = \frac{w_{uv}}{\deg u},\]
with the remark that this probability is zero whenever $E(u,v) = \emptyset$. The random walk transition matrix is the matrix $M$ given by  $M_{v,u} = \frac{w_{uv}}{\deg u}$.

If $G$ is a strongly connected aperiodic graph, then the Perron--Frobenius Theorem tells us there is a unique positive vector $\varphi = (\varphi(u))_{u\in V}$ with $||\varphi||_1 = 1$ such that $M\varphi = \varphi$ (see \cite[Proposition~1.14 and Theorem~4.9]{2006/Levin--Peres}). This vector $\varphi$ is called the \textbf{stationary distribution} of $G$. Moreover, for any starting distribution $\psi$ on the vertices of $G$, we have $\lim_{n\to\infty} M^n\psi = \varphi$.\footnote{If we drop the connectivity hypothesis, then $\varphi$ is neither positive nor unique. Meanwhile, a periodic graph will still have a stationary distribution, but convergence to it is not granted.}

When $G$ is an \emph{undirected} graph,
the stationary distribution is the vector $\varphi$ where
\[
    \varphi(u) = \frac{\deg u}{2|E|}
    \quad
    \text{for}
    \quad 
    u \in V
    \,;
\]
we see immediately that this is indeed the stationary distribution,
because
\[
    \varphi(u) 
    = 
    \frac{\deg u}{2|E|} 
    = 
    \sum_{v\in N(u)} \frac{1}{\deg v} \frac{\deg v}{2|E|}
    \,.
\]
However, when $G$ is a \emph{directed} graph, there is no closed-form formula for the stationary distribution of the random walk. Even the principal ratio $\frac{\max_{u\in V}\varphi(u)}{\min_{u\in V}\varphi(u)}$ of the distribution can be difficult to bound, and it can be exponentially large even when degree bounds such as $\delta \leq \deg u \leq \Delta$, for all $u\in V$, are known \cite{2016/Aksoy--Chung--Peng}.

\subsection{Directed graphs and linear imbalance}

The following definition tries to restrict the amount of allowed ``directedness'' in a graph, so that we are able to find closed-form stationary distributions for isogeny graphs. It applies directly to the graph $\SSIG[2][2][11]$ displayed in Figure~\ref{fig:p-11}.

\begin{definition}
Let $G = (V, E, w)$ be a directed weighted graph. We say $G$ has \textbf{linear imbalance} if there exists a vertex partition $V=A_1 \sqcup \cdots \sqcup A_n$ and a bijection 
\[
    E(u,v) \overset{\dualof{(\cdot)}}{\to} E(v,u)
\]
for each pair of adjacent vertices $u, v \in V$, such that
\begin{enumerate}
    \item If $u,v\in A_i$, then for each $e\in E(u,v)$, $w(e) = w(\dualof{e})$.
    \item For each $i\neq j$ there exists a rational number $m_{ij}$, such that if $u\in A_i$, $v\in A_j$, and $e\in E(u,v)$, then $w(e) = m_{ij}\cdot w(\dualof{e})$.
\end{enumerate}
In particular $m_{ji} = m_{ij}^{-1}$, and we can set $m_{ii} = 1$.
\end{definition}

We can see $G$ as an undirected graph if we forget the weights, due to the existence of the bijections $E(u,v) \overset{\dualof{(\cdot)}}{\to} E(v,u)$. However, the presence of weights changes the definition of the random walk on G, and in particular the stationary distribution will be different. We now want to compute this distribution.

\begin{proposition}\label{proposition:linear-imbalance-distribution}
Let $G = (V, E)$ be a linear imbalance graph with partition $V = A_1 \sqcup \cdots \sqcup A_n$. 
Assume all vertices of each given class $A_i$ have the same degree $d_i$, i.e., $\deg(u) = d_i$ for all $u\in A_i$.

Suppose there exists a non-zero solution
$(\alpha_1,\dots,\alpha_n)$ to the system
of equations
\begin{equation}\label{eqs2}
    \frac{m_{ji}}{d_j} \alpha_j = \frac{1}{d_i} \alpha_i
    \quad
    \text{for every }
    i, j 
    \text{ such that } E(A_i,A_j) \neq \emptyset
    \,.
\end{equation}
Define the vectors $\tilde\varphi = (\tilde\varphi(u))_{u\in V}$ by $\tilde\varphi(u) = \alpha_i$ if $u\in A_i$, and $\varphi = \tilde\varphi / ||\tilde\varphi||_1$.

The vector $\varphi$ is a stationary distribution for the random walk on $G$. Moreover, the random walk on $G$ is a reversible Markov chain.
\end{proposition}
\begin{proof}
We need to check that
\[\tilde \varphi(u) = \sum_{v\in N(u),\ e\in E(u,v)} \frac{w(\dualof{e})}{\deg v} \tilde \varphi(v).\]
Say $u \in A_i$, and label its neighbors $v_1,\dots,v_{t_u}$ (inside the classes $A_{j_1},\dots,A_{j_{t_u}}$). Then the previous equation becomes
\[
    \tilde \varphi(u) 
    = \sum_{v\in N(u),\ e\in E(u,v)} \frac{w(\dualof{e})}{\deg v} \tilde \varphi(v)
    = \sum_{k=1}^{t_u} \frac{m_{j_k i} w_{u v_k}}{d_{j_k}} \tilde\varphi({v_k}).
\]
Substituting the values of $\tilde\varphi(u)$ and $\tilde\varphi(v_k)$, we get the equation
\[
    \alpha_i 
    = \sum_{k=1}^{t_u} \frac{m_{j_k i} w_{u v_k}}{d_{j_k}} \alpha_{j_k}.
\]
Using Equations~\eqref{eqs2}, we get
\[\alpha_i = \sum_{k=1}^{t_u} \frac{w_{u v_k}}{d_i} \alpha_i
=\left(\sum_{k=1}^{t_u} w_{u v_k}\right)\frac{1}{d_i}\alpha_i,\]
which is trivially true.

We say a Markov chain is reversible if, for all states $u,v$, we have
\[\varphi(u) P(u,v) = \varphi(v)P(v,u)\]
where $P(u,v)$ is the probability of walking from $u$ to $v$. In our case, this equation becomes
\[\alpha_i \frac{w_{u v}}{d_i} = \alpha_j \frac{w_{v u}}{d_j}\]
whenever $u\in A_i, v\in A_j$, which is always satisfied (after dividing both sides by $w_{v u}$). This proves the reversibility of the chain.
\end{proof}

Proposition~\ref{proposition:linear-imbalance-distribution} imposes a total of $\binom{n}{2}$ equations, which may or may not yield a solution. However, we can reduce the number of necessary equations if the graph is connected and has {composable} linear imbalance.

\begin{definition}
Let $G$ and $A_i$ be as above. Construct an undirected graph $\mathcal G = (\mathcal V, \mathcal E)$ with vertices $\mathcal V = \{a_1, \dots, a_n\}$ and with edges $\mathcal E = \{\{a_i,a_j\} \mid E(A_i,A_j)\neq\emptyset \}$. We say $G$ has \textbf{composable linear imbalance}\footnote{This is also known in the Markov chain literature as the Kolmogorov criterion, and it characterises chain reversibility. We use this term as it provides more meaning to our setting.}
if for any two neighboring vertices $a_i, a_j$ and for any path in $\mathcal G$ (with distinct edges and vertices)  $a_i = a_{i_0} \to a_{i_1} \to \cdots \to a_{i_k} = a_j$ from $a_i$ to $a_j$ we have
\[m_{ji} = m_{ji_{k-1}} m_{i_{k-1}i_{k-2}}\cdots m_{i_1i}.\]
\end{definition}

Every undirected graph has composable linear imbalance by defining any partition on its set of vertices. Or, alternatively, a linear imbalance graph is undirected if and only if $m_{ij} = 1$ for all $i,j$.

\begin{lemma}\label{lemma:equation-reducing}
    Let $G = (V, E)$ be a connected graph satisfying the same conditions as in 
    Proposition~\ref{proposition:linear-imbalance-distribution}. 
    If $G$ has composable linear imbalance, 
    then the set of equations
\begin{equation}
    \frac{m_{ji}}{d_j} \alpha_j = \frac{1}{d_i} \alpha_i
\end{equation}
can be reduced to a set of $n - 1$ equations, where $n$ is the number of classes in the vertex partition of $G$.
\end{lemma}
\begin{proof}
Recall $V = A_1 \sqcup \cdots \sqcup A_n$, and let $\mathcal G$ be the graph associated to this partition. Let $\mathcal T$ be any spanning tree of $\mathcal G$. 

Consider the system of $n - 1$ equations 
\(\frac{m_{ji}}{d_j} \alpha_j = \frac{1}{d_i} \alpha_i\)
whenever $\{a_i, a_j\}$ is an edge in $\mathcal T$. We claim this system is equivalent to the full system. Indeed, for any two vertices $a_i, a_j \in \mathcal T$ such that $E(A_i,A_j) \neq \emptyset$, let 
\[a_i = a_{i_0} \to a_{i_1} \to \cdots \to a_{i_k} = a_j\]
be a path in $\mathcal T$ from $a_i$ to $a_j$. Using the newly defined system, we get the equation
\[\frac{1}{d_i}\alpha_i = \frac{m_{ji_{k-1}} m_{i_{k-1}i_{k-2}}\cdots m_{i_1i}}{d_j}\alpha_j,\]
which by composability gives us the desired equation
\(\frac{1}{d_i}\alpha_i = \frac{m_{ji}}{d_j}\alpha_j.\)
\end{proof}

\begin{example}
\begin{enumerate}
    \item This result can be illustrated by computing the stationary distribution
    for the random walk over $\SSIG[1]$ with $p\equiv 11\pmod{12}$ (the
    other possibilities for $p$ are special cases of this). We partition the set of vertices $V$ into three sets, $A_0 = V \setminus \{\EC_0,\EC_{1728}\}$, $A_1 = \{\EC_0\}$, and $A_2 = \{\EC_{1728}\}$. This partition gives the graph composable linear imbalance, with $m_{01} = 3$, $m_{02} = 2$, and $m_{12} = 2/3$. The graph $\mathcal G$ is a triangle\footnote{It is actually a tree in many cases, but the computation is the same.}, which imposes three linear equations in three variables, but we get a spanning tree $\mathcal T$ by removing any edge. For instance, we get the equations
\[\frac{1}{\ell + 1} \alpha_0 = \frac{3}{\ell + 1} \alpha_1
        \quad \text{and}\quad 
\frac{1}{\ell + 1} \alpha_0 = \frac{2}{\ell + 1} \alpha_2\]
which are satisfied by $(\alpha_0, \alpha_1, \alpha_2) = (1,1/3,1/2)$.
\item The same procedure can be applied to the graph $\SSIG[2][2][11]$ displayed in Figure~\ref{fig:p-11}. We have a disjoint partition in five one-vertex sets, and the multipliers $m_{ij}$ between them are given by ratios of sizes of automorphism groups. By the same procedure as above, the stationary distribution is given by the vector
\[
    (\alpha_{\AV_1}, \alpha_{\AV_2}, \alpha_{\EC_{1728}^2},\alpha_{\EC_0^2},\alpha_\Pi) = \frac{144}{121}\cdot\left(\frac{1}{12}, \frac{1}{6}, \frac{1}{16},\frac{1}{36}, \frac{1}{12}\right).
\]
\end{enumerate}
\end{example}

\begin{corollary}\label{corollary:stationary-distribution}
Let $G = (V,E)$ be a connected linear imbalance graph with a vertex partition $V=A_1 \sqcup \cdots \sqcup A_n$. Suppose that for each $1 \leq i \leq n$ there exists a positive real number $g_i$ such that for all $i, j$, $m_{ij} = \frac{g_i}{g_j}$. Then $G$ has composable linear imbalance, and it has stationary distribution $\varphi = \tilde \varphi/||\tilde \varphi||_1$, where
\[
    \tilde \varphi(u) = \frac{d_i}{g_i} = \frac{\deg(u)}{g_i}
    \quad
    \text{whenever}
    \quad
    u \in A_i
    \,.
\]
\end{corollary}
\begin{proof}
The fact that $G$ has composable linear imbalance is trivial from the equalities $m_{ij}= \frac{g_i}{g_j}$. From Lemma \ref{lemma:equation-reducing}, the equations
\(\frac{m_{ji}}{d_j} \alpha_j = \frac{1}{d_i} \alpha_i\)
are satisfied for all $i,j$ with $E(A_i,A_j)\neq \emptyset$. But these equations correspond to
\(\frac{g_j}{d_j} \alpha_j = \frac{g_i}{d_i} \alpha_i\)
which are trivially satisfied by setting $\alpha_i = d_i/g_i$.
\end{proof}

We discuss now the mixing rate of a graph $G$ satisfying the hypotheses
of the last result. Let $M_G$ be the random walk matrix. We define an inner product on $\RR^{|V(G)|}$, denoted by $\langle\cdot,\cdot\rangle_\varphi$, by
\[\langle f,g\rangle_\varphi = \sum_{u\in V(G)} f(u)g(u)\varphi(u)\,.\]

\begin{lemma}[\cite{2006/Levin--Peres}, Lemma~12.2]
The reversible property of the random walk on G implies:
\begin{enumerate}
    \item The inner product space
        $(\RR^{|V(G)|},\langle\cdot,\cdot\rangle_\varphi)$ has an
        orthonormal basis $\{f_j : 1 \le j \le |V(G)|\}$ 
        of real-valued left eigenvectors of $M_G$, 
        corresponding to real eigenvalues $\{\lambda_j: 1 \le j \le |V(G)|\}$.
    \item Given a random walk $u = u_0 \to \cdots \to u_n \to \cdots$, for all $v\in V(G)$ we have
    \begin{equation}\label{eq:matrix-decomposition}
        \frac{\Prob{u_n = v}}{\varphi(v)} 
        = 
        1 + \sum_{j=2}^{|V(G)|} f_j(u)f_j(v)\lambda_j^n
        \,.
    \end{equation}
\end{enumerate}
\end{lemma}

In particular, if the graph $G$ is connected and aperiodic, then we know
\[1 = \lambda_1 > \lambda_2 \geq \cdots \geq \lambda_{|V(G)|} > -1.\]
Letting $\lambda_\star(G) = \max\{|\lambda| \mid \lambda \text{ is an eigenvalue of } M_G, \lambda\neq 1\}$, we have the following result bounding the mixing rate of the random walk.

\begin{proposition}\label{proposition:mixing-rate}
Consider a random walk $u = u_0 \to \cdots \to u_n \to \cdots$, and let $v\in V(G)$ be any vertex. If $u \in A_i$ and $v \in A_j$, we have
\[
    \left|\Prob{u_n = v} - \varphi(v)\right|
    \leq
    \lambda_\star(G)^n \sqrt{\frac{\deg(v)}{\deg (u)}\frac{g_i}{g_j}}.
\]
\end{proposition}
\begin{proof}
    We adapt the proof of \cite[Theorem~12.3]{2006/Levin--Peres}. Using
    Eq.~\eqref{eq:matrix-decomposition} and the Cauchy--Schwarz inequality we get
    \begin{align*}
        \left|\frac{\Prob{u_n = v}}{\varphi(v)} - 1\right| 
        & \leq
        \sum_{j=2}^{|V(G)|} |f_j(u)f_j(v)|\lambda_\star(G)^n 
        \\
        & \leq 
        \lambda_\star(G)^n \left( \sum_{j=2}^{|V(G)|} f_j^2(u) \sum_{j=2}^{|V(G)|}f_j^2(v) \right)^{1/2}
        \,.
    \end{align*}
Let $\delta_w$ be the function
\[\delta_w(u) = 
\begin{cases}
1 &\text{if }w = u,\\
0 &\text{if }w \neq u.
\end{cases}\]
This function can be written in the following way, using the orthonormal basis of functions $\{f_j\}_{j=1}^{|V(G)|}$:
\[\delta_w = \sum_{j=1}^{|V(G)|} \langle \delta_w, f_j\rangle_\varphi f_j
= \sum_{j=1}^{|V(G)|} f_j(w) \varphi(w) f_j.\]
From this we obtain
\begin{align*}
    \varphi(w) 
    = \langle \delta_w,\delta_w \rangle_\varphi
    &= \left\langle \sum_{j=1}^{|V(G)|} f_j(w) \varphi(w) f_j , \sum_{j=1}^{|V(G)|} f_j(w) \varphi(w) f_j \right\rangle_\varphi\\
    &= \varphi(w)^2 \sum_{j=1}^{|V(G)|} f_j^2(w),
\end{align*}
which implies $\sum_{j=2}^{|V(G)|} f_j^2(w) < \varphi(w)^{-1}$. Combining this with the first stated inequality we get
\[
    |\Prob{u_n = v} - \varphi(v)| \leq \lambda_\star(G)^n \sqrt{\frac{\varphi(v)}{\varphi(u)}}
    \,;
\]
the result follows on substituting the values of $\varphi$ obtained in Corollary~\ref{corollary:stationary-distribution}.
\end{proof}

Proposition~\ref{proposition:mixing-rate} is the analog of classical results on random walk mixing in undirected graphs: \cite[Theorem~5.1]{1993/Lovasz} for the general case,  \cite[Theorem~3.3]{2006/Hoory--Linial--Widgerson} for regular graphs, and \cite{2012/Menares} and \cite[Theorem~1]{2016/Galbraith--Petit--Silva} for supersingular isogeny graphs.

\subsection{Isogeny graphs as linear imbalance graphs}
\label{sec:isogeny-graphs-distribution}

Our results so far allow us to give the stationary distribution and
convergence rate for superspecial isogeny graphs. But we can state a much
more general result, and apply the same theory to interesting isogeny subgraphs.

\begin{theorem}\label{theorem:isogeny-graphs-distribution}
Let $G$ be a finite, connected and aperiodic subgraph of $\IG$, such that for each edge $\classof{\phi}$ in $G$, its dual edge $\classof{\dualof{\phi}}$ is also in $G$. 
    \begin{enumerate}
        \item
            The stationary distribution of the random walk in $G$ is given by $\varphi_G = \tilde\varphi_G / ||\tilde\varphi_G||_1$, 
            \[
                \tilde\varphi_G({\AV}) = \frac{\deg(\AV)}{\#RA(\AV)}
                \,,
            \]
            where $\deg(\AV)$ denotes the number of isogenies in $G$ with domain $\AV$.
        \item
            The mixing rate is $\lambda_\star(G)$. More precisely, if
            $\AV_0 \to \cdots \to \AV_n \to \cdots$ is a random walk, and $\AV$
            is any vertex of $G$, then the convergence to the stationary distribution is given by
            \begin{equation}\label{eq:convergence}
                |\operatorname{Pr}[\AV_n \cong \AV] - \varphi_G(\AV)| 
                \leq
                \lambda_\star(G)^n \sqrt{\frac{\deg\AV}{\deg\AV_0} \frac{\#\RAut(\AV_0)}{\#\RAut(\AV)}}
                \,.
            \end{equation}
    \end{enumerate}
\end{theorem}
\begin{proof}
    For Part (1): 
    Lemma~\ref{lemma:key} tells us that $G$ has linear imbalance, by
    partitioning its set of vertices according to the reduced
    automorphism group of each variety. Indeed, for
    any two neighbouring PPAVs $\AV$ and $\AV'$ in \IG, 
    we have 
\[
    \frac{w_{\AV,\AV'}}{w_{\AV',\AV}}
    =
    \frac{\#\RAut(\AV)}{\#\RAut(\AV')}
    \,.
\]
We can refine this partition further so that all nodes in a single class
    have the same degree. This way, all hypotheses of
    Proposition~\ref{proposition:linear-imbalance-distribution} and
    Corollary~\ref{corollary:stationary-distribution} are satisfied,
    yielding the stated distribution.
    Part (2) then follows from Proposition~\ref{proposition:mixing-rate}.
\end{proof}

Theorem~\ref{theorem:isogeny-graphs-distribution} is true for all
superspecial isogeny graphs $\SSIG$, as they are connected and
non-bipartite \cite[Corollary~18]{2020/Jordan--Zaytman} and hence
aperiodic. In fact, we can always produce a loop if $g$ is even:
if 
\(
    \phi\colon \EC \to \EC'
\)
is an elliptic \(\ell\)-isogeny,
then the product $(\ell,\dots,\ell)$-isogeny
\begin{equation}\label{eq:elliptic-loop}
    (\EC \times \EC')^{g/2}
    \xrightarrow{\phi\times\dualof{\phi}\times \cdots \times \phi\times\dualof{\phi}}
    (\EC \times \EC')^{g/2}
\end{equation}
is a loop in \SSIG.
If $g$ is odd, we let $\psi_1\colon \EC \to \EC$, $\psi_2\colon \EC \to
\EC$ be two elliptic curve isogenies of respective degrees $\ell^e$ and
$\ell^f$ with $e$ and $f$ coprime (this exists, since $\SSIG[1]$ is non-bipartite \cite[Corollary~18]{2020/Jordan--Zaytman} and so aperiodic). Then, by constructing the previous isogeny $\phi\times\dualof{\phi}\times \cdots \times \phi\times\dualof{\phi}$ in genus $g-1$, we get two isogenies
\begin{align*}
    (\phi\times \dualof{\phi})^e \times \cdots \times (\phi\times\dualof{\phi})^e \times \psi_1,\\
    (\phi\times \dualof{\phi})^f \times \cdots \times (\phi\times\dualof{\phi})^f \times \psi_2,
\end{align*}
where exponentiation means composition ($\phi \times \dualof{\phi}$ is an endomorphism of $\EC\times\EC'$), representing two cycles of coprime lengths $e$ and $f$ in $\SSIG$.

\subsection{Bounds on eigenvalues}
\label{sec:eigenvalue-bounds}

If we fix $g$ and $\ell$, and we have a constant
$\lambda=\lambda(g,\ell) < 1$ such that $\lambda_\star(\SSIG) \leq
\lambda$ for all $p$, then we get a family of graphs with good expansion
properties\footnote{Note that they should not be called
\textit{expander} graphs: this term is reserved for regular undirected
graphs.}. Combining this with Equation~\eqref{eq:convergence}, we conclude that the diameter of each graph is $O(\log p)$, a property that also holds for regular expander graphs.

Given a $d$-regular undirected graph $G$ with $\lambda_\star(G)$ as second largest eigenvalue (in absolute value), we have $d\cdot \lambda_\star(G) \geq 2\sqrt{d - 1} - o_n(1)$. Here $o_n(1)$ is a quantity that tends to zero for fixed $d$ when the number of vertices $n$ goes to infinity. 
If $d\cdot \lambda_\star(G) \leq 2\sqrt{d - 1}$, then \(G\) is said to be \textbf{Ramanujan} \cite{2006/Hoory--Linial--Widgerson}. Ramanujan graphs have optimal expansion properties. 

Isogeny graphs of supersingular elliptic curves are Ramanujan \cite{1990/Pizer}, and it was hoped that this property would extend to the more general graphs $\SSIG$ \cite[Hypothesis~1]{2020/Costello--Smith}. We have shown $\SSIG$ does not fit into the definition of an \textit{expander graph} for $g\geq 2$, due to the presence of non-trivial reduced automorphism groups. However, we may still ask for bounds on $\lambda_\star(\SSIG)$, as a Ramanujan property of sorts. Now, letting $N_g(\ell)$ be the out-degree of the vertices in $\SSIG$, we ask a question: for which $g$, $\ell$ and $p$, if any, does the bound
\[
    N_g(\ell)\cdot \lambda_\star(\SSIG) \leq 2\sqrt{N_g(\ell) - 1}
\]
hold? 

Jordan and Zaytman \cite{2020/Jordan--Zaytman} have given a first counterexample: $\SSIG[2][2][11]$ is not Ramanujan, as the second largest eigenvalue of the adjacency matrix is $7 + \sqrt{3}$, which is larger than $2\sqrt{N_2(2) - 1} = 2\sqrt{15 - 1}$. 

We have gathered evidence that the same behaviour also occurs for (at least) all graphs $\SSIG[2][2][p]$ for primes $11 \leq p \leq 601$. For all these primes, the superspecial Richelot isogeny graph fails to be Ramanujan, and in fact most values of $\lambda_\star$ (except for a few small primes) are very close to $11.5/15$. Giving a theoretical reason for this behaviour is left as future work.

The eigenvalues and diameters of each graph can be found in Appendix~\ref{sec:data}. In Section~\ref{sec:connectivity} we prove that both the subgraph of Jacobians and the subgraph of elliptic products satisfy the hypotheses to have convergence to a stationary distribution, and so our data also includes their eigenvalues and diameters.

We now refine the previously stated conjectures on superspecial graphs.

\begin{conjecture}
\label{conjecture:eigenvalue-bound}
    For all $g$ and $\ell$, 
    there exists a fixed $\lambda = \lambda(g,\ell) <1$ such that
    \[
        \lambda_\star(\SSIG) \leq \lambda
        \quad
        \text{for every prime }
        p \ge 5
        \,.
    \]
    In the case \(g = 2\) and \(\ell = 2\), 
    we conjecture that
    \[
        \frac{11}{15} \le \lambda_\star(\SSIG[2][2]) \le \frac{12}{15}
        \quad
        \text{for every prime }
        p \geq 41
        \,.
    \] 
\end{conjecture}

%% file: ppas.tex
\section{
    The Richelot isogeny graph
}
\label{sec:Richelot-graph}
\label{sec:ppas}

From now on, we focus on the case \(g = 2\) and \(\ell = 2\).
Richelot~\cite{1836/Richelot,1837/Richelot} 
gave the first explicit construction for \((2,2)\)-isogenies, 
so the \((2,2)\)-isogeny graph of principally polarized abelian surfaces
(PPASes) is called the \emph{Richelot isogeny graph}.

Let \(\AV_{0}\) be a PPAS with full rational 2-torsion.
There are 15 rational Lagrangian subgroups
\(K_{1},\ldots,K_{15}\) of \(\AV_{0}[2]\),
and each is the kernel of a rational \((2,2)\)-isogeny
\[
    \phi_i: \AV_{0} \rightarrow \AV_i := \AV_{0}/K_i
    \,.
\]
This means that every vertex in the \((2,2)\)-isogeny graph has
out-degree 15.
In general, none of the isogenies or codomains are isomorphic.
The algorithmic construction of the isogenies and codomains
depends fundamentally on whether \(\AV_{0}\) is a Jacobian or an
elliptic product.
We recall the Jacobian case 
in~\S\ref{sec:Richelot},
and the elliptic product case
in~\S\ref{sec:elliptic-product-isogenies}.

Before going further,
we recall the explicit classification
of (reduced) automorphism groups of PPASes.
In contrast with elliptic curves,
where (up to isomorphism) only two curves
have nontrivial reduced automorphism group,
with PPASes we see much richer structures
involving many more vertices in \(\IG[2][2]\).

\subsection{Jacobians of genus-2 curves}
\label{sec:RAut-Jac}

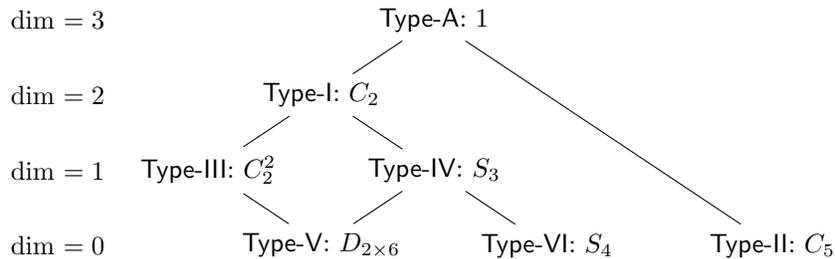
\begin{figure}[ht]
    \centering
    \begin{tikzpicture}
        \node (TypeA) at (0,0) {\TypeA: \(1\)} ;

        \node (TypeII) at (4.5,-3) {\TypeII: \(C_5\)} ;

        \node (TypeI) at (-1.5,-1) {\TypeI: \(C_2\)} ;

        \node (TypeIII) at (-3,-2) {\TypeIII: \(C_{2}^2\)} ;

        \node (TypeIV) at (0,-2) {\TypeIV: \(S_3\)} ;

        \node (TypeV) at (-1.5,-3) {\TypeV: \(D_{2\times 6}\)} ;

        \node (TypeVI) at (1.5,-3) {\TypeVI: \(S_4\)} ;

        \path (TypeII) edge (TypeA)  ;
        \path (TypeI) edge (TypeA)   ;
        \path (TypeIII) edge (TypeI) ;
        \path (TypeIV) edge (TypeI)  ;
        \path (TypeV) edge (TypeIII) ;
        \path (TypeV) edge (TypeIV)  ;
        \path (TypeVI) edge (TypeIV) ;

        \node (Mzero) at (-5,-3) {dim \(= 0\)} ;
        \node (Mone) at (-5,-2) {dim \(= 1\)} ;
        \node (Mtwo) at (-5,-1) {dim \(= 2\)} ;
        \node (Mthree) at (-5,0) {dim \(=3\)} ;
    \end{tikzpicture}
    \caption{%
        The taxonomy of reduced automorphism groups for genus-2 Jacobians.
        Dimensions on the left are of the loci on each level 
        in the 3-dimensional moduli space of PPASes.
        Lines connect sub-types and super-types; 
        specialization moves down the page.
    }
    \label{fig:RA-Jacobian}
\end{figure}

Bolza~\cite{1887/Bolza} has shown that there are seven possible reduced automorphism groups
for Jacobian surfaces
(provided \(p > 5\)).
Figure~\ref{fig:RA-Jacobian} 
gives Bolza's taxonomy, defining names (``types'')
for each of the reduced automorphism groups.

We can identify the isomorphism class of a Jacobian \(\Jac{\XC}\)
using the Clebsch invariants \(A\), \(B\), \(C\), \(D\)
of \(\XC\),
which are homogeneous polynomials of degree 2, 4, 6, and 10
in the coefficients of the sextic defining \(\XC\).
Detailed formul\ae{} appear in~\S\ref{sec:Clebsch}.

\subsection{Products of elliptic curves}
\label{subsection:products-of-elliptic-curves}

Elliptic products always have nontrivial reduced automorphism groups,
because
\(\RAut(\EC\times\EC')\) always contains the involution
\[
    \sigma := [1]_{\EC}\times[-1]_{\EC'}
    \,.
\]
Note that \(\sigma\) fixes every Lagrangian subgroup of \((\EC\times\EC')[2]\)
(though this is not true for \((\EC\times\EC')[\ell]\) if \(\ell > 2\)),
so \(\sigma\) always has an impact on the Richelot isogeny graph.

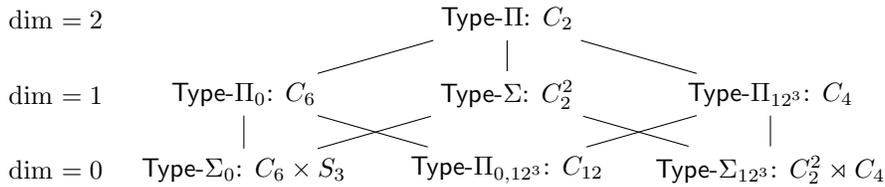
\begin{figure}[ht]
    \centering
    \begin{tikzpicture}

        \node (TypeExE) at (0,0) {\TypeExE: \(C_2\)} ;

        \node (TypeE2) at (0,-1) {\TypeEsquared: \(C_2^2\)} ;

        \node (TypeExE0) at (-3.5,-1) {\TypeExEzero: \(C_6\)} ;

        \node (TypeExE1728) at (3.5,-1) {\TypeExEtwelvecubed: \(C_4\)} ;

        \node (TypeE02) at (-3.5,-2) {\TypeEzerosquared: \(C_6\times S_3\)} ;

        \node (TypeE0xE1728) at (0,-2) {\TypeEzeroxEtwelvecubed: \(C_{12}\)} ;

        \node (TypeE17282) at (3.5,-2) {\TypeEtwelvecubedsquared: \(C_2^2\rtimes C_4\)} ;
                 
        \path (TypeExE0) edge (TypeExE)  ;
        \path (TypeE2) edge (TypeExE)   ;
        \path (TypeExE1728) edge (TypeExE) ;
        \path (TypeE02) edge (TypeExE0)  ;
        \path (TypeE02) edge (TypeE2)  ;
        \path (TypeE17282) edge (TypeExE1728)  ;
        \path (TypeE17282) edge (TypeE2)  ;
        \path (TypeE0xE1728) edge (TypeExE0)  ;
        \path (TypeE0xE1728) edge (TypeExE1728)  ;

        \node (Mzero) at (-6,-2) {dim \(= 0\)} ;
        \node (Mone) at (-6,-1) {dim \(= 1\)} ;
        \node (Mtwo) at (-6,0) {dim \(= 2\)} ;
    \end{tikzpicture}
    \caption{%
        The taxonomy of reduced automorphism groups of elliptic products.
        Dimensions on the left are of the loci on each level
        in the 3-dimensional moduli space of PPASes.
        Lines connect sub-types and super-types; 
        specialization moves down the page.
    }
    \label{fig:RA-elliptic}
\end{figure}

Proposition~\ref{prop:elliptic-product-RAuts}
shows that there are
seven possible reduced automorphism groups for elliptic product surfaces
(provided \(p > 3\)),
and
Figure~\ref{fig:RA-elliptic}
gives a taxonomy of reduced automorphism groups
analogous to that of Figure~\ref{fig:RA-Jacobian}.
We identify the isomorphism class of
an elliptic product \(\EC\times\EC'\)
using the \(j\)-invariants \(j(\EC)\) and \(j(\EC')\)
(an unordered pair when \(\EC\not\cong\EC'\),
and a single \(j\)-invariant when \(\EC\cong\EC'\)).

\begin{proposition}
    \label{prop:elliptic-product-RAuts}
    If \(\AV\) is an elliptic product surface,
    then (provided \(p > 3\)) 
    there are seven possibilities for the isomorphism type of \(\RAut(\AV)\).
    \begin{enumerate}
        \item
            If \(\AV \cong \EC\times\EC'\)
            for some \(\EC \not\cong \EC'\),
            then
            one of the following holds:
            \begin{itemize}
                \item \TypeExE:
                    \(\{j(\EC),j(\EC')\}\cap\{0,1728\} = \emptyset\),
                    and \(\RAut(\AV) \cong C_2\).
                \item \TypeExEzero:
                    \(j(\EC) = 0\) or \(j(\EC') = 0\),
                    and \(\RAut(\AV) \cong C_6\).
                \item \TypeExEtwelvecubed:
                    \(j(\EC) = 1728\) or \(j(\EC') = 1728\),
                    and \(\RAut(\AV) \cong C_4\).
                \item \TypeEzeroxEtwelvecubed:
                    \(\{j(\EC),j(\EC')\} = \{0,1728\}\),
                    and \(\RAut(\AV) \cong C_{12}\).
            \end{itemize}
        \item
            If \(\AV \cong \EC^2\)
            for some \(\EC\),
            then one of the following holds:
            \begin{itemize}
                \item \TypeEsquared:
                    \(j(\EC) \notin\{0,1728\}\),
                    and
                    $\RAut(\AV)\cong C_2^2$.
                \item \TypeEzerosquared:
                    \(j(\EC) = 0\),
                    and
                    $\RAut(\AV)\cong C_6\times S_3$.
                \item \TypeEtwelvecubedsquared:
                    \(j(\EC) = 1728\),
                    and
                    $\RAut(\AV)\cong C_2^2\rtimes C_4$.
            \end{itemize}
    \end{enumerate}
\end{proposition}
\begin{proof}
    Recall that if \(\EC\) is an elliptic curve,
    then:
    if \(j(\EC) = 0\)
    then \(\Aut(\EC) = \subgrp{\rho} \cong C_6\);
    if \(j(\EC) = 1728\)
    then \(\Aut(\EC) = \subgrp{\iota} \cong C_4\);
    and otherwise
    \(\Aut(\EC) = \subgrp{[-1]} \cong C_2\).

    For Part~(1):
    if $\EC\not\cong \EC'$, 
    then $\Aut(\EC\times \EC') \cong \Aut(\EC)\times \Aut(\EC')$. 
    If $\Aut(\EC) = \langle \alpha\rangle$
    and $\Aut(\EC') = \langle \beta\rangle$,
    then $\Aut(\EC\times \EC') =\langle \alpha\times [1], [1]\times \beta\rangle$.
    Notice that $\beta^d = [-1]$ for $d=1, 2$ or $3$, 
    so if $j(\EC)\notin\{0,1728\}$, 
    then $\RAut(\EC\times \EC') \cong \Aut(\EC')$,
    which proves the first three cases.
    For the remaining \TypeEzeroxEtwelvecubed case, 
    the automorphism $[\rho]\times[\iota]$ has exact order $12$, 
    proving $\RAut(\EC\times\EC') \cong C_{12}$.

    For Part~(2): in this case $\Aut(\EC^2)$ certainly
    contains $\Aut(\EC)^2$ as a subgroup,
    but we also have the involution
    \(
        \tau \colon (P,Q) \mapsto (Q,P)
    \).
    The existence of $\tau$ makes \(\Aut(\EC^2)\) non-abelian, 
    because $(\beta\times \gamma) \circ \tau = \tau \circ (\gamma\times \beta)$
    for any $\beta,\gamma\in \Aut(\EC)$.
    If $\Aut(\EC) = \langle\alpha\rangle$, then $\Aut(\EC^2) = \langle\alpha\times[1], [1]\times\alpha,\tau\rangle$
    is the wreath product \(\Aut(\EC)\wr \subgrp{\tau}\),
    i.e., the semidirect
    product $(\Aut(\EC)\times \Aut(\EC))\rtimes \subgrp{\tau}$. 
    More explicitly: if \(\Aut(\EC) = \subgrp{\alpha}\),
    then
    \[
        \Aut(\EC^2) \cong \langle a, b, \tau 
        \mid 
        a^d = b^d = \tau^2 = 1,\ 
        ab=ba,\ 
        a\tau = \tau b \rangle,
    \]
    where $a = \alpha\times[1]$, $b = [1]\times\alpha$, and $d\in\{2,4,6\}$ is the order of $\alpha$. 
    Taking the quotient by \([-1]_{\EC^2}\),
    we identify the reduced automorphism groups
    using GAP's \texttt{IdGroup}~\cite{GAP4}.
\end{proof}

\subsection{Implications for isogeny graphs}

The vertices in \IG corresponding to PPAVs with nontrivial reduced
automorphism groups form interesting and inter-related structures.
We highlight a few of these facts for \(g = 2\) and \(\ell = 2\).

Katsura and Takashima observe that if we take a Jacobian vertex
\(\classof{\Jac{\XC}}\) in \IG[2][2], 
then the number of elliptic-product neighbours of 
\(\classof{\Jac{\XC}}\)
is equal to the number of involutions \(\alpha\) in \(\RAut(\Jac{\XC})\)
induced by involutions in \(\Aut(\Jac{\XC})\)
(see~\cite[Proposition 6.1]{2020/Katsura--Takashima}).
In particular: general \TypeA vertices and the unique \TypeII vertex
have \emph{no} elliptic product neighbours;
\TypeI and \TypeIV vertices, and the unique \TypeVI vertex,
have \emph{one} elliptic product neighbour;
and the \TypeIII vertices and the unique \TypeV vertex
have \emph{two} elliptic-square neighbours.
By explicit computation of Richelot isogenies
we can (slightly) extend Katsura and Takashima's results
to give the complete description of weighted edges with codomain types
for each of the vertex types in Table~\ref{tab:atlas-summary}.
The inter-relation of reduced automorphism groups
and neighbourhoods of vertices and edges in the Richelot isogeny graph
is further investigated (and illustrated) 
in~\cite{2020/Florit--Smith:atlas}.

\begin{table}[htp]
    \centering
    \begin{tabular}{cccl|cccl}
        Vertex & \#Edges & \(w\) & Neighbour 
        &
        Vertex & \#Edges & \(w\) & Neighbour 
        \\
        \hline
        \hline
        \TypeA                     & 15 & 1 & \TypeA
        &
        \multirow{2}{*}{\TypeExE}  & 9 & 1 & \TypeExE
        \\
        \cline{1-4} 
        \multirow{3}{*}{\TypeI}    & 1  & 1 & \TypeExE
        &
        \phantom{\TypeExE}         & 6 & 1 & \TypeI 
        \\
        \cline{5-8}
        \phantom{\TypeI}           & 6  & 1 & \TypeI
        &
        \multirow{2}{*}{\TypeExEzero} & 3 & 3 & \TypeExE
        \\
        \phantom{\TypeI}           & 4  & 2 & \TypeA
        &
        \phantom{\TypeExEzero}        & 2 & 3 & \TypeI
        \\
        \cline{1-4} 
        \cline{5-8}
        \TypeII                    & 3  & 5 & \TypeA
        &
        \multirow{3}{*}{\TypeExEtwelvecubed} & 3 & 1 & \TypeExEtwelvecubed
        \\
        \cline{1-4} 
        \multirow{4}{*}{\TypeIII}  & 1  & 1 & (loop)
        &
        \phantom{\TypeExEtwelvecubed}        & 3 & 2 & \TypeExE
        \\
        \phantom{\TypeIII}         & 2  & 1 & \TypeEsquared
        &
        \phantom{\TypeExEtwelvecubed}        & 3 & 2 & \TypeI
        \\
        \cline{5-8}
        \phantom{\TypeIII}         & 4  & 2 & \TypeI
        &
        \multirow{3}{*}{\TypeEzeroxEtwelvecubed} & 1 & 3 & \TypeExEtwelvecubed
        \\
        \phantom{\TypeIII}         & 1  & 4 & \TypeA
        &
        \phantom{\TypeEzeroxEtwelvecubed}        & 1 & 6 & \TypeExE
        \\
        \cline{1-4} 
        \multirow{3}{*}{\TypeIV}   & 1 & 3 & \TypeExE
        &
        \phantom{\TypeEzeroxEtwelvecubed}        & 1 & 6 & \TypeI
        \\
        \cline{5-8}
        \phantom{\TypeIV}          & 3 & 3 & \TypeI
        &
        \multirow{5}{*}{\TypeEsquared} & 1 & 1 & (loop)
        \\
        \phantom{\TypeIV}          & 3 & 1 & \TypeIV
        &
        \phantom{\TypeEsquared}                   & 3 & 2 & \TypeExE
        \\
        \cline{1-4} 
        \multirow{5}{*}{\TypeV}                   & 1 & 3 & (loop)
        &
        \phantom{\TypeEsquared}                   & 3 & 1 & \TypeEsquared
        \\
        \phantom{\TypeV}                          & 1 & 1 & \TypeEzerosquared 
        &
        \phantom{\TypeEsquared}                   & 1 & 2 & \TypeI
        \\
        \phantom{\TypeV}                          & 1 & 3 & \TypeEsquared
        &
        \phantom{\TypeEsquared}                   & 3 & 1 & \TypeIII
        \\
        \cline{5-8}
        \phantom{\TypeV}                          & 1 & 6 & \TypeI
        &
        \multirow{3}{*}{\TypeEzerosquared}        & 1 & 3 & (loop)
        \\
        \phantom{\TypeV}                          & 1 & 2 & \TypeIV
        &
        \phantom{\TypeEzerosquared}               & 1 & 9 & \TypeEsquared
        \\
        \cline{1-4} 
        \multirow{3}{*}{\TypeVI}                  & 1 & 1 & (loop)
        &
        \phantom{\TypeEzerosquared}               & 1 & 3 & \TypeV 
        \\
        \cline{5-8}
        \phantom{\TypeVI}                         & 1 & 6 & \TypeEsquared
        &
        \multirow{4}{*}{\TypeEtwelvecubedsquared} & 1 & 3 & (loop)
        \\
        \phantom{\TypeVI}                         & 2 & 4 & \TypeIV 
        &
        \phantom{\TypeEtwelvecubedsquared}        & 1 & 4 & \TypeEsquared
        \\
        \cline{1-4} 
        & & & 
        &
        \phantom{\TypeEtwelvecubedsquared}        & 1 & 4 & \TypeExEtwelvecubed
        \\
        & & & 
        &
        \phantom{\TypeEtwelvecubedsquared}        & 1 & 4 & \TypeIII
        \\
        \cline{5-8}
    \end{tabular}
    \caption{%
        Number of edges, weights, and types of neighbours for
        vertices in \IG[2][2] by reduced automorphism type.
        Observe that the edge numbers multiplied by their weights
        always sum to 15.
        Neighbour types may change under specialization
        (or for particular values of~\(p\)),
        acquiring reduced automorphisms.
        See~\cite{2020/Florit--Smith:atlas} for details.
    }
    \label{tab:atlas-summary}
\end{table}

\begin{remark}
    Each \TypeIV vertex 
    has a triple edge to an elliptic-product neighbour.
    In fact,
    the factors of the product
    are always \(3\)-isogenous
    (cf.~\cite[\S3]{2001/Gaudry--Schost}).
    The unique \TypeVI vertex
    is a specialization of \TypeIV,
    and in this case the \TypeExE neighbour
    specializes to the square of an elliptic curve
    with \(j\)-invariant 8000 (which has an endomorphism of degree 3).
    The unique \TypeV vertex
    is also a specialization of \TypeIV,
    and
    in this case
    the \TypeExE neighbour specializes to
    the square of an elliptic curve of \(j\)-invariant 54000
    (which as an endomorphism of degree 3);
    one of the \TypeIV neighbours degenerates to the square of an
    elliptic-curve with \(j\)-invariant 0,
    while
    the other two merge, yielding a weight-2 edge;
    and one of the \TypeI neighbours specializes
    to the \TypeV vertex, yielding a loop,
    while the other two merge,
    yielding a weight-6 edge.
\end{remark}

\begin{remark}
    Every \TypeIII vertex (and the unique \TypeV vertex)
    has \emph{two} elliptic-square neighbours:
    these are the squares of a pair of 
    2-isogenous elliptic curves~\cite[\S4]{2001/Gaudry--Schost}.
    In this way,
    \TypeIII vertices in \IG[2][2] correspond to undirected edges 
    (i.e., edges modulo dualization of isogenies) in \IG[1][2].
\end{remark}

Ibukiyama, Katsura, and Oort have computed the precise number of superspecial
genus-2 Jacobians (up to isomorphism) of each reduced automorphism
type~\cite[Theorem 3.3]{1986/Ibukiyama--Katsura--Oort}.
We reproduce their results for \(p > 5\)
in Table~\ref{tab:number-of-superspecial-vertices},
completing them with the number of superspecial elliptic products
of each automorphism type
(which can be easily derived from the well-known formula
for the number of supersingular elliptic curves over~\(\FF_{p^2}\)).

\begin{table}[ht]
    \centering
    \begin{tabular}{|r|l||r|l|}
        \hline
        Type & Vertices in \(\RichelotIG\)
        &
        Type & Vertices in \(\RichelotIG\)
        \\
        \hline
        \hline
        \multirow{2}{*}{\TypeI}
               & \( \frac{1}{48}(p-1)(p-17) \)
        &       
        \TypeExE & \( \frac{1}{2}N_p(N_p - 1) \)
        \\
        \cline{3-4}
        {}
               & \qquad \quad \(
                   + \frac{1}{4}\epsilon_{1,p}
                   + \epsilon_{2,p}
                   + \epsilon_{3,p}
                 \)
        &
        \TypeExEzero & \( \epsilon_{3,p}N_p \)
        \\
        \hline
        \TypeII & \(\epsilon_{5,p}\)
        &
        \TypeExEtwelvecubed & \( \epsilon_{1,p}N_p \)
        \\
        \hline
        \TypeIII & \( 
                      \frac{3}{2}N_p 
                      + \frac{1}{2}\epsilon_{1,p} 
                      - \frac{1}{2}\epsilon_{2,p} 
                      - \frac{1}{2}\epsilon_{3,p}
                   \)
        &
        \TypeEzeroxEtwelvecubed & \(\epsilon_{1,p}\cdot\epsilon_{3,p}\)
        \\
        \hline
        \TypeIV & \(2N_p + \epsilon_{1,p} - \epsilon_{2,p}\)
        &
        \TypeEsquared & \(N_p\) 
        \\
        \hline
        \TypeV & \(\epsilon_{3,p}\)
        &
        \TypeEzerosquared & \(\epsilon_{3,p}\)
        \\
        \hline
        \TypeVI & \(\epsilon_{2,p}\)
        &
        \TypeEtwelvecubedsquared & \(\epsilon_{1,p}\)
        \\
        \hline
        \TypeA & \multicolumn{3}{l|}{\(
                   \frac{1}{2880}(p-1)(p^2-35p+346)
                   - \frac{1}{16}\epsilon_{1,p}
                   - \frac{1}{4}\epsilon_{2,p}
                   - \frac{2}{9}\epsilon_{3,p}
                   - \frac{1}{5}\epsilon_{5,p}
                 \)}
        \\
        \hline
    \end{tabular}
    \caption{The number of vertices in \(\SSIG[2]\)
        of each reduced automorphism type.
        Here
        \(\epsilon_{1,p} = 1\) if \(p \equiv 3\pmod{4}\), 0 otherwise;
        \(\epsilon_{2,p} = 1\) if \(p \equiv 5,7\pmod{8}\), 0 otherwise;
        \(\epsilon_{3,p} = 1\) if \(p \equiv 2\pmod{3}\), 0 otherwise;
        \(\epsilon_{5,p} = 1\) if \(p \equiv 4 \pmod{5}\), 0 otherwise;
        and \(N_p = (p-1)/12 - \epsilon_{1,p}/2 - \epsilon_{3,p}/3\)
        is the number of supersingular elliptic curves over \(\FF_{p^2}\) 
        with reduced automorphism group \(C_2\).
    }
    \label{tab:number-of-superspecial-vertices}
\end{table}


%% file: random-richelot.tex
\section{
    Random walks in the superspecial Richelot isogeny graph
}
\label{sec:random-Richelot}

We now specialize the results of~\S\ref{sec:random-walks} to the case
\(g = 2\), \(\ell = 2\),
and consider some cryptographic applications.

\subsection{Random walks}
Given an isogeny graph \(G\) 
satisfying the hypotheses of Theorem~\ref{theorem:isogeny-graphs-distribution},
we let 
\[
    K_G = \max_{\AV, \AV_0} \sqrt{
        \frac{\deg_G \AV}{\deg_G \AV_0}\frac{\#\RAut(\AV_0)}{\#\RAut(\AV)}
    }.
\]
If we put $G=\SSIG[2][2]$ and consider the reduced automorphism groups
in Proposition~\ref{prop:elliptic-product-RAuts}, then $K_G = 6$.
Together with Conjecture~\ref{conjecture:eigenvalue-bound},
this gives us precise constants for the convergence of the random walk distribution on the Richelot isogeny graph. We will say that a vector $\psi\in \RR^{|V(G)|}$ approximates the stationary distribution $\varphi$ of the graph $G$ with an error of $\varepsilon>0$ if for each vertex $u\in V(G)$, $|\psi(u) - \varphi(u)|\leq \varepsilon$. A random walk of length $n$ approximates the stationary distribution with error $\varepsilon$ if the distribution given by the walk at step $n$ does so.

\begin{theorem}
    \label{theorem:walk-length}
    Assume Conjecture~\ref{conjecture:eigenvalue-bound} for \(g = 2\)
    and \(\ell = 2\):
    that is, assume that $\lambda_\star(\SSIG[2][2])\leq \frac{12}{15}$ for all $p\geq 41$.
    A random walk of length $n\geq 4.5m\log p + 9$ approximates the stationary distribution on $\SSIG[2][2]$ with an error of $\frac{1}{p^m}$. In particular, a random walk of length
    \[
        n\geq 18\log p + 9
    \]
    approximates the stationary distribution with an error of $\frac{1}{p^4}$.
\end{theorem}
\begin{proof}
    Set $G=\SSIG[2][2]$. Given a random walk $\AV_0 \to \cdots \to \AV_n \to \cdots$ and a vertex $\AV$, then for all $n$ we have
    \begin{equation*}
        |\operatorname{Pr}[\AV_n \cong \AV] - \varphi_G(\AV)| 
        \leq
        \lambda_\star(G)^n \sqrt{\frac{\deg_G\AV}{\deg_G\AV_0} \frac{\#\RAut(\AV_0)}{\#\RAut(\AV)}}
        \leq 6\lambda_\star(G)^n.
    \end{equation*}
    The inequality $6\lambda_\star(G)^n \leq \frac{1}{p^m}$ is satisfied as long as 
    \[
        n\geq \frac{m\log p + \log 6}{\log(\lambda_\star(G)^{-1})}.
    \]
    Since $\log 6 / \log(15/12) \leq 9$ and $1/\log(15/12)\leq 4.5$, if
    \(
        n \geq 4.5m\log p + 9
    \)
    then the above inequalities are satisfied. The particular case of $m=4$ follows.
\end{proof}

\subsection{Distributions of subgraphs}
\label{sec:distributions-of-subgraphs}

If we perform a random walk on $\SSIG[2][2]$, 
 we will encounter a certain number of products of elliptic curves along the way.
We can try to predict the ratio of elliptic products to visited nodes:
a first guess could be that this ratio matches 
the proportion of such nodes in the entire
graph, which is asymptotic to $\frac{10}{p}$
(see~\cite[Proposition~2]{2020/Castryck--Decru--Smith}). 

However, this is not the empirical proportion that we observe in
our experiment, which consists in performing $10,000$ random walk
steps in $\SSIG[2][2]$ and counting the number $N$ of elliptic products
encountered in our path. The ratio $N/10,000$ of elliptic products to visited nodes
is closer to $\frac{5}{p}$, as seen in Table~\ref{tab:random-walks-products}.

\begin{table}[ht]
\centering
\begin{tabular}{l|l|l|l|l|l|l}
    $p$                     & 101      & 307      & 503     & 701      & 907     & 1103     \\ \hline
$N$ & 415      & 201      & 130     & 64       & 50      & 44       \\
    Ratio                   & \(4.1915/p\) & \(6.1707/p\) & \(6.539/p\)
    & \(4.4864/p\) & \(4.535/p\) & \(4.8532/p\)
\end{tabular}
\caption{Number of elliptic products encountered in a $10,000$-step random walk for several primes. The third row shows the proportion scaled relative to each prime.}
\label{tab:random-walks-products}
\end{table}

Theorem~\ref{theorem:isogeny-graphs-distribution},
in combination with the classification of reduced automorphism groups in Proposition~\ref{prop:elliptic-product-RAuts},
gives us the true proportion of elliptic product nodes in random walks.
We have $\frac{p^3}{2880} + O(p^2)$ Jacobians with trivial reduced
automorphism group (this is the picture for ``almost all'' nodes in the
graph: only $O(p^2)$ have nontrivial reduced automorphisms), and there
are $\frac{p^2}{288} + O(p)$ elliptic products. However, all but $O(p)$
of those products have a reduced automorphism group of order 2,
confirming that the (asymptotic) expected proportion of elliptic
products in a random walk is equal to $\frac{1}{2}\times\frac{10}{p} = \frac{5}{p}$. Similarly, we could compute proportions for each abelian surface type given in Section~\ref{sec:ppas}.

If we combine this with the conjectured upper bound for
$\lambda_\star(\SSIG[2][2])$, then we can give the interpretation that elliptic products are evenly distributed in the graph, in the sense that any node is within very few steps of an elliptic product (much less than diametral distance).

\subsection{The superspecial isogeny problem in genus \(2\) and beyond}
The general problem of constructing an isogeny between two superspecial
\(g\)-dimensional PPAVs \(\AV_g\) and \(\AV_g'\) over \(\FF_{p^2}\)
was studied in~\cite{2020/Costello--Smith}.
The algorithm proceeds
by computing isogenies
\(\phi: \AV_g \to \AV_{g-1}\times\EC\)
and
\(\phi': \AV_g' \to \AV_{g-1}'\times\EC'\)
where \(\AV_{g-1}\) 
and \(\AV_{g-1}'\) 
have dimension \(g-1\)
and \(\EC\) and \(\EC'\) are elliptic curves,
before computing an elliptic isogeny \(\EC \to \EC'\)
and (recursively) computing an isogeny \(\AV_{g-1} \to \AV_{g-1}\),
then combining the results to produce an isogeny \(\AV_{g} \to \AV_{g}'\).
The key step is computing the isogenies \(\phi\) and \(\phi'\)
to product PPAVs.
The expected complexity of this step is heuristic,
and assumes that the isogeny graph of superspecial PPAVs
has good expansion properties to ensure that \(O(p)\) 
isogeny walks of length \(O(\log p)\) will result in
a walk to a product variety with probability \(O(1)\).
Of course,
in practice one cannot simply take walks of length \(O(\log p)\):
we need a proper bound on the length of these walks
(essentially, we need the constant hidden by the big O).

Our results show if we admit
Conjecture~\ref{conjecture:eigenvalue-bound},
then 
the expected complexity of the algorithm in~\cite{2020/Costello--Smith}
is rigorous for \(g = 2\),
and we can bound the required walk lengths using the claimed eigenvalue bounds
as in Theorem~\ref{theorem:walk-length}.
In particular,
for \(g = 2\) and \(\ell = 2\),
it suffices to use walks of length \(26\log_2(p) + 8\).

\subsection{Richelot isogeny hash functions}
Recall the Richelot-isogeny hash function
of~\cite{2020/Castryck--Decru--Smith},
which is based on walks in \(\RichelotIG\).
A binary representation of the data to be hashed 
is broken into a series of three-bit chunks;
each of the eight possible three-bit values corresponds to the choice of a step in
\(\RichelotIG\) such that the composition of the prior step with the
current step is a \((4,4)\)-isogeny.
The hash value is (derived from) the invariants of the final vertex in the walk.

Our results show that finding an input \(m\) driving a walk into
the induced subgraph \(\SSIG[2][2]^E\) on the elliptic product vertices
would immediately yield collisions in the hash function.
Indeed, looking at Table~\ref{tab:atlas-summary},
we see that every vertex in \(\SSIG[2][2]^E\)
has either outgoing edges with multiplicity greater than~\(1\),
or a \TypeI neighbour with outgoing edges with multiplicity greater
than~\(1\).
This means that there are multiple kernels,
and thus multiple \(3\)-bit input chunks,
that produce steps to the same neighbour;
in this way, given a walk to \(\SSIG[2][2]^E\),
with at most two further steps
we can construct explicit hash collisions.

Since the forward steps in these walks are restricted to a subset of eight of the
fourteen possible onward edges at each vertex,
the results in~\S\ref{sec:eigenvalue-bounds}
do not apply directly here.
Still, they give us reason to
hope that these restricted random walks will approximate the uniform
distribution on \(\RichelotIG\) very quickly.
If adversaries can compute walks into \(\SSIG[2][2]^E\)
after an expected \(O(p)\) steps,
as they can with unrestricted walks,
then 
they can use walks into \(\SSIG[2][2]^E\)
to construct hash collisions
in an expected \(\softO(p)\) operations,
which is exponentially fewer
than the \(O(p^{3/2})\) required by generic attacks.

\subsection{Genus 2 SIDH analogues}
Our results also have constructive cryptographic applications.
For example,
consider  
the genus-2 SIDH
analogue proposed by Flynn and Ti~\cite{2019/Flynn--Ti},
a postquantum key exchange algorithm
based on commuting random walks in \(\SSIG[2][2]\)
and \(\SSIG[2][3]\).
The walks involved are very short---on the order of
\(\frac{1}{2}\log_2p\) steps each---and much shorter
than the bound of Theorem~\ref{theorem:walk-length}.
Our results therefore
imply that this genus-2 SIDH analogue is overwhelmingly unlikely to encounter
\(\SSIG[2][\ell]^E\),
provided the base vertex is chosen sensibly.

%% file: connectivity.tex
\section{Connectivity and diameters}\label{sec:connectivity}

We mentioned in~\S\ref{sec:random-walks}
that Theorem~\ref{theorem:isogeny-graphs-distribution} can be applied to
study distributions in interesting isogeny subgraphs of the superspecial isogeny graph.
Let us then distinguish three subgraphs of $\SSIG$, each taken to be the
induced subgraph defined by its set of vertices:
\begin{itemize}
    \item $\SSIG^J$, the subgraph of Jacobians;
    \item $\SSIG^P$, the subgraph of reducible PPAVs (product
        varieties); and
    \item $\SSIG^E$, the subgraph of products of elliptic curves.
\end{itemize}
(Observe that $\SSIG[2]^P = \SSIG[2]^E$).
Understanding the connectivity of such subgraphs can be useful both
when analysing the algorithms that work with them,
and when studying the distribution of vertices in the full supersingular graph. 

\begin{proposition}
    \label{prop:SSIGP-SSIGE-aperiodic} 
    The graphs $\SSIG^P$ and $\SSIG^E$ are connected and aperiodic
    for all \(g\), \(\ell\), and \(p\). In
    particular, both graphs satisfy the hypotheses of Theorem~\ref{theorem:isogeny-graphs-distribution}.
\end{proposition}
\begin{proof}
    It is enough to see that $\SSIG^E$ is connected and aperiodic, 
    since it is a subgraph of $\SSIG^P$ and given a product variety we can find a product isogeny to an elliptic product by the connectivity of $\SSIG$.
    We obtain connectivity from the fact that $\SSIG^E$ has a spanning subgraph which is a quotient of the tensor product of $g$ copies of the supersingular isogeny graph $\SSIG[1]$. Since $\SSIG[1]$ is aperiodic, it contains an odd cycle and so $(\SSIG[1])^{\otimes g}$ is connected \cite{1962/Weichsel}.
    We have already proved aperiodicity, since
    in~\S\ref{sec:isogeny-graphs-distribution} we constructed loops and paths of coprime lengths in $\SSIG^E$.
\end{proof}

Proposition~\ref{prop:SSIGP-SSIGE-aperiodic} generalizes immediately to
any connected component of the general graph $\IG$ that contains elliptic products. 

Conjecture~2 of~\cite{2020/Castryck--Decru--Smith}
proposes that the subgraph of the superspecial Richelot isogeny graph
supported on the Jacobians is connected;
Theorem~\ref{theorem:Jacobian-connectivity}
confirms and proves this conjecture.
(We should be able to give a similar statement for the Jacobian subgraph even without the superspecial condition,
but the technique that we use only allows us to prove it for the case $g=2$, $\ell = 2$.)

\begin{theorem}
    \label{theorem:Jacobian-connectivity}
    The graph of Jacobians $\SSIG[2][2]^J$ is connected and aperiodic.
    In particular, it satisfies the hypotheses of Theorem~\ref{theorem:isogeny-graphs-distribution}.
\end{theorem}
\begin{proof}
    To see $\SSIG[2][2]^J$ is connected, it is enough to check that the
    subgraph containing all \TypeI Jacobians is connected. Indeed, any
    two Jacobians $J_1$ and $J_2$ are connected by a path in
    $\SSIG[2][2]$, and we only need to ensure that subpaths between
    \TypeI Jacobians can be modified to avoid elliptic products. This is
    always possible by Lemma~\ref{lemma:eliminate-ExE} below.

    The aperiodicity for primes $p \geq 13$ comes from the fact that
    there are always \TypeIII Jacobians, which always have a $(2,2)$-endomorphism. One checks easily that $\SSIG[2][2]^J$ has at least one loop when $p$ is $7$ or $11$. Indeed, for $p=7$ the unique \TypeVI vertex has a $(2,2)$-endomorphism $\phi$ with weight $w(\classof{\phi}) = 9$, while for $p=7$ the unique \TypeV vertex has a $(2,2)$-endomorphism $\psi$ with $w(\classof{\psi}) = 3$.
\end{proof}

\begin{lemma}
    \label{lemma:eliminate-ExE}
Given a path $\classof{J_0} \to \classof{\EC\times \EC'} \to \classof{\AV}$ in $\Gamma_2(2;p)$, where $J_0$ is a Jacobian, $\EC\times \EC'$ is an elliptic product, and $\AV$ is any PPAS, there exists either:
\begin{enumerate}
    \item A length-2 path 
    \[\classof{J_0} \to \classof{J_1} \to \classof{\AV},\]
    where $J_1$ is a Jacobian, if the original path represents a $(4,2,2)$-isogeny, or
    
    \item A length-4 path
    \[\classof{J_0} \to \classof{J_1} \to \classof{J_2} \to \classof{J_3} \to \classof{\AV},\]
    where each $J_i$ is a Jacobian, if the original walk represents a $(4,4)$-isogeny.
\end{enumerate}
\end{lemma}

\begin{proof}
    Case 1. The original path represents a $(4,2,2)$-isogeny, $\phi$.
    Up to isomorphism, 
    $\phi$ factors into a
    composition of two $(2,2)$-isogenies in 3 ways:
    \begin{itemize}
        \item $\phi: J_0 \to \EC\times \EC' \to \AV$, 
        \item $\phi_1: J_0 \to \AV_1 \to \AV$, and
        \item $\phi_2: J_0 \to \AV_2 \to \AV$.
    \end{itemize}
    The isogenies $J_0 \to \AV_i$ each have one nontrivial kernel point in
    common with $J_0 \to \EC\times \EC'$.  We know that $\classof{J_0}$
    has at most two elliptic-product neighbours 
    (see Table~\ref{tab:atlas-summary}).
    Recall the language of \emph{quadratic splittings} detailed in
    Appendix~\ref{sec:Richelot}:
    the Lagrangian subgroups of \(J_0[2]\)
    correspond to factorizations of \(f(x)\) into three coprime
    quadratics,
    where \(C_0: y^2 = f(x)\) is a sextic model for the genus-2 curve
    generating \(J_0\),
    and the codomain of the corresponding \((2,2)\)-isogeny is an
    elliptic product precisely when
    the three quadratics are linearly dependent.
    After a coordinate transformation, we can suppose that $J_0 \to \EC\times \EC'$ is a Richelot isogeny with $\ker(J_0 \to \EC\times \EC') = \{ x^2 - a^2, x^2 - b^2, x^2 - c^2 \}$. Relabelling $(a,b,c)$ if necessary, we can assume the point common to $\ker(J_0 \to \EC\times \EC')$, $\ker(J_0 \to \AV_1)$, and $\ker(J_0 \to \AV_2)$ corresponds to $x^2 - a^2$, and thus
    \begin{align*}
        \ker(J_0 \to \AV_1) 
        & = 
        \{x^2 - a^2, x^2 - (b+c)x + bc, x^2 + (b+c)x + bc\}
        \shortintertext{and}
        \ker(J_0 \to \AV_2) 
        & =
        \{x^2 - a^2, x^2 - (b-c)x - bc, x^2 + (b-c)x - bc\}
        \,.
    \end{align*}
    It is easy to check that the determinants of these two triples cannot both vanish unless the original curve is singular.

Case 2. The original walk represents a $(4,4)$-isogeny, $\phi$.
We can always choose a neighbour $\classof{J_2} \neq \classof{J_0}$ of $\classof{\EC\times \EC'}$ such that $J_0 \to \EC\times \EC' \to J_2$ and $J_2 \to \EC\times \EC' \to \AV$ both 
represent $(4,2,2)$-isogenies. Now apply Case 1 to each of these,
eliminating $\EC\times \EC'$ from the middle of each length-2 path, and compose the results.
\end{proof}

\begin{remark}
When $J_0$ is \TypeIII or \TypeV in the $(4,2,2)$-isogeny case, it is
    possible that we obtain $\classof{J_0} = \classof{J_1}$, so we
    actually simplify to a length-1 path $\classof{J_0} \to
    \classof{\AV}$.
    Further,
    in the $(4,4)$-isogeny case, 
    we can even have
    $\classof{J_0} = \classof{J_2}$,
    and then we can simplify the original length-2 path (and the modified
    length-4 one) to the length-1 path $\classof{J_0}\to\classof{\AV}$.
\end{remark}

\begin{corollary}
    \label{cor:diameter}
    The diameters of $\SSIG[2][2]$ and $\SSIG[2][2]^J$ satisfy
    \[
        \operatorname{diam}(\SSIG[2][2]) - 2
        \leq
        \operatorname{diam}(\SSIG[2][2]^J) 
        \leq 
        2\operatorname{diam}(\SSIG[2][2])
        \,.
    \]
\end{corollary}
\begin{proof}
The first inequality comes from the fact that every elliptic product has
    a Richelot isogeny to a Jacobian. For the second one, apply
    Lemma~\ref{lemma:eliminate-ExE} repeatedly to bound the distance between any two nodes in $\SSIG[2][2]^J$.
\end{proof}

The lower bound of Corollary~\ref{cor:diameter} is tight, as seen for $\SSIG[2][2][521]$.
Our experimental results suggest that the upper bound has some room for improvement.

%% file: example-p-47.tex
\section{An example: the superspecial Richelot graph for \(p = 47\)}

We now exemplify our results on the Richelot isogeny graph for $p=47$.
The graph $\SSIG[2][2][47]$ has an appropriate size to observe
interesting behaviour.
In particular, since $p\equiv 11\bmod 12$ and $p\equiv 2\bmod 5$,
all of the vertex types 
described in Section~\ref{sec:Richelot-graph}
except $\TypeII$ appear.
Table~\ref{tab:p47-counts} lists the exact counts for each vertex type.

\begin{table}[ht]
    \centering
    \begin{tabular}{|r||c|c|c|c|c|c|c|c|c|c|c|c|c|c|}
    \hline
        Type \(T\) & A  & I  & II & III & IV & V  & VI & $\Sigma$ & $\Pi$ & $\Pi_{12^3}$ & $\Pi_0$  & $\Sigma_{12^3}$ & $\Pi_{0,12^3}$ & $\Sigma_0$  \\
    \hline\hline
        \(\#A_T\)  & 14 & 31 & 0  & 4   & 6  & 1  & 1  & 3        & 3     & 3            & 3        & 1               & 1              & 1 \\
    \hline
        \(g_T\)    & 1  & 2  & -- & 4   & 6  & 12 & 23 & 4        & 2     & 4            & 6        & 16              & 12             & 36 \\
    \hline
    \end{tabular}
    \caption{%
        Vertex counts for each type in the graph $\SSIG[2][2][47]$.
        Here \(A_T\) denotes the subset of vertices of type \(T\),
        while \(g_T\) is the corresponding value of \(g_i\) in
        Corollary~\ref{corollary:stationary-distribution}.%
    }
    \label{tab:p47-counts}
\end{table}

\begin{figure}
    \centering
    \includegraphics[width=\textwidth]{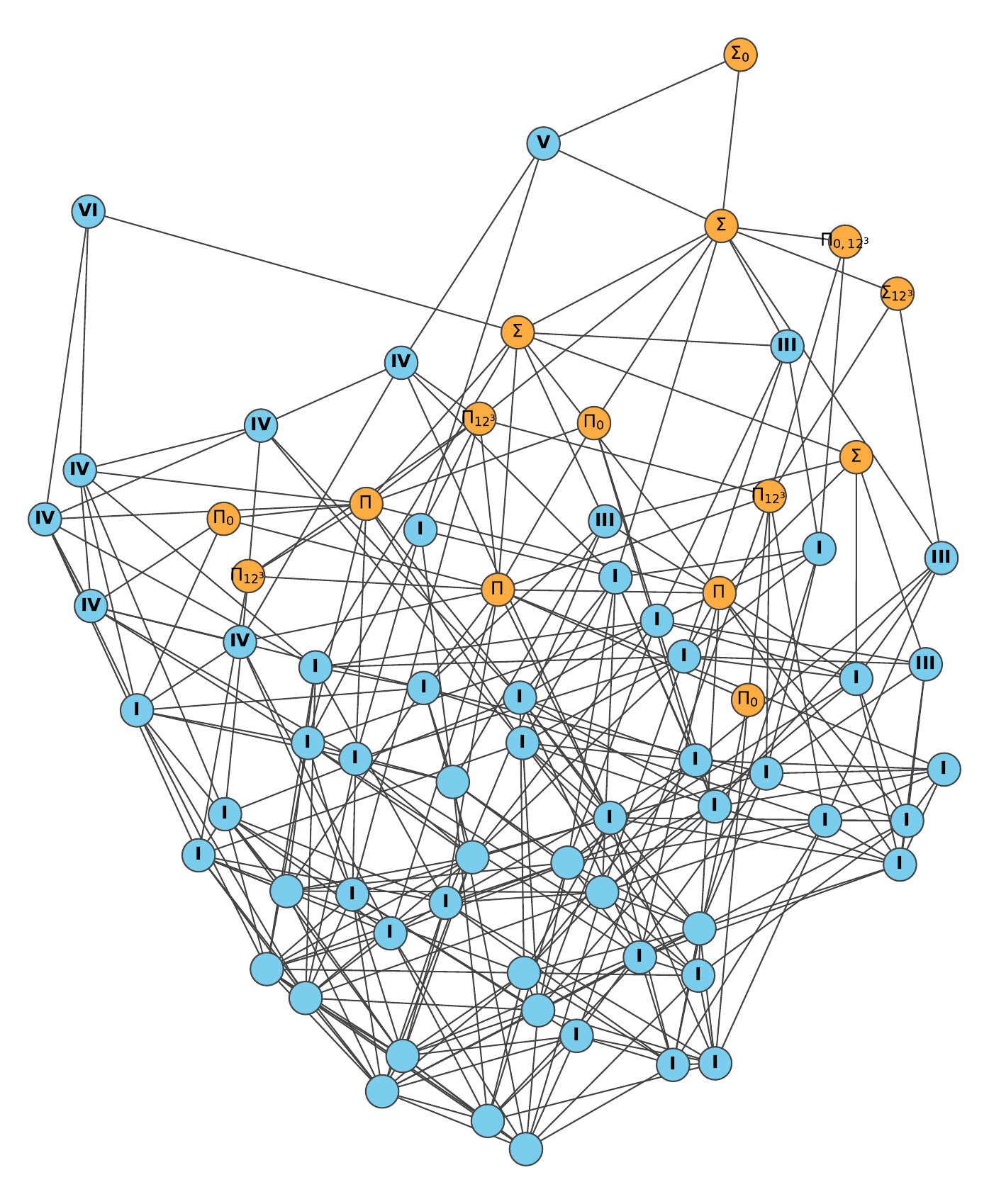}
    \caption{The superspecial Richelot isogeny graph for $p=47$.
    Vertices are labeled with their types; unlabeled vertices are
    $\TypeA$, with trivial reduced automorphism group. Loops are omitted.}
    \label{fig:p47}
\end{figure}

Let us compute the stationary distribution for the full graph $\SSIG[2][2][47]$.
First, we partition the vertex set according to each type: $A_{\TypeA}$
contains the 14 $\TypeA$ vertices, $A_{\TypeI}$ the 31 $\TypeI$ vertices, and so on.
In the notation of Corollary~\ref{corollary:stationary-distribution},
if \(A_i = A_T\) for a type \(T\),
then the values of \(g_i\) are the \(g_T\) in
Table~\ref{tab:p47-counts}.
%
(In general, we would also have $g_{II} = 1/5$.)
Since all vertices have 15 Lagrangian subgroups in their two-torsion,
Corollary~\ref{corollary:stationary-distribution} says that
(after normalization) the stationary distribution is given by
\[
    \tilde\varphi(\AV) = \frac{1}{g_T}
    \qquad
    \text{whenever $\AV$ is of type $T$}.
\]
We can observe this partially in Figure~\ref{fig:p47}. The picture lacks
the edge weights, which we have omitted for the sake of clarity.
Nevertheless, we see clearly that vertices with larger reduced
automorphism groups are more isolated, because lots of isogenies are
identified through automorphisms.
This makes these vertices harder to reach in a random walk,
so they have a smaller value in the stationary distribution.

We may also compute the stationary distributions of the subgraphs $\SSIG[2][2][47]^J$ and $\SSIG[2][2][47]^E$. Recall from Table~\ref{tab:atlas-summary} that the degrees in these graphs are no longer regular: for example, a $\TypeA$ varieties have 15 isogenies to other Jacobians, while $\TypeI$ varieties have 14 isogenies to other Jacobians and a single isogeny to a product of elliptic curves. The stationary probability for a vertex $\AV$ of type $T$ is
\[
    \tilde\varphi(\AV) = \frac{\deg\AV}{g_T}
    \qquad
    \text{whenever $\AV$ is of type $T$},
\]
where $\deg\AV$ is now the number of isogenies from $\AV$ to vertices
\emph{in the same graph}, and $g_T$ is defined as above. 

\begin{figure}%
    \centering
    \subfloat[\centering Jacobian subgraph]{{\includegraphics[width=5cm]{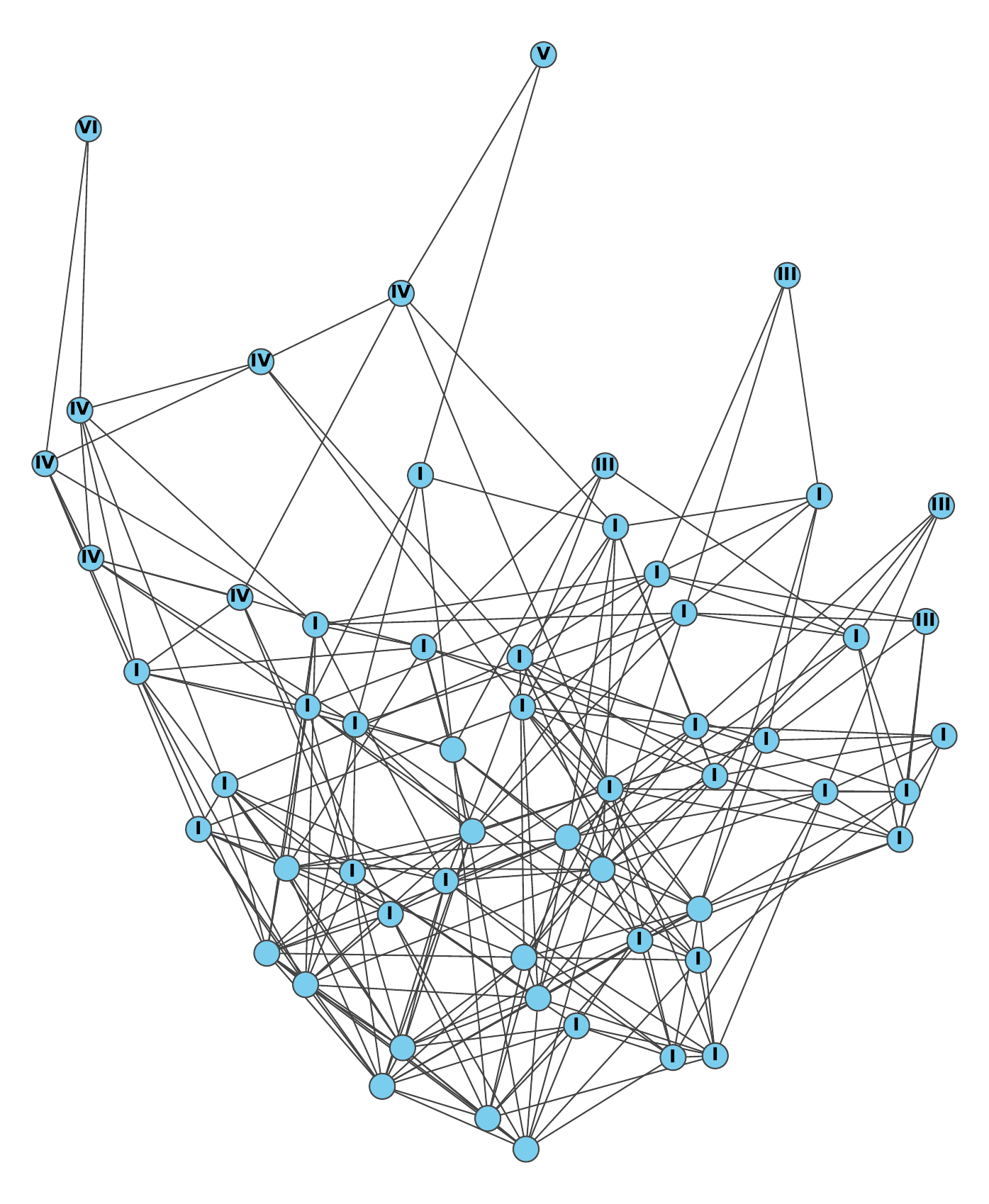} }}%
    \qquad
    \subfloat[\centering Elliptic product subgraph]{{\includegraphics[width=5cm]{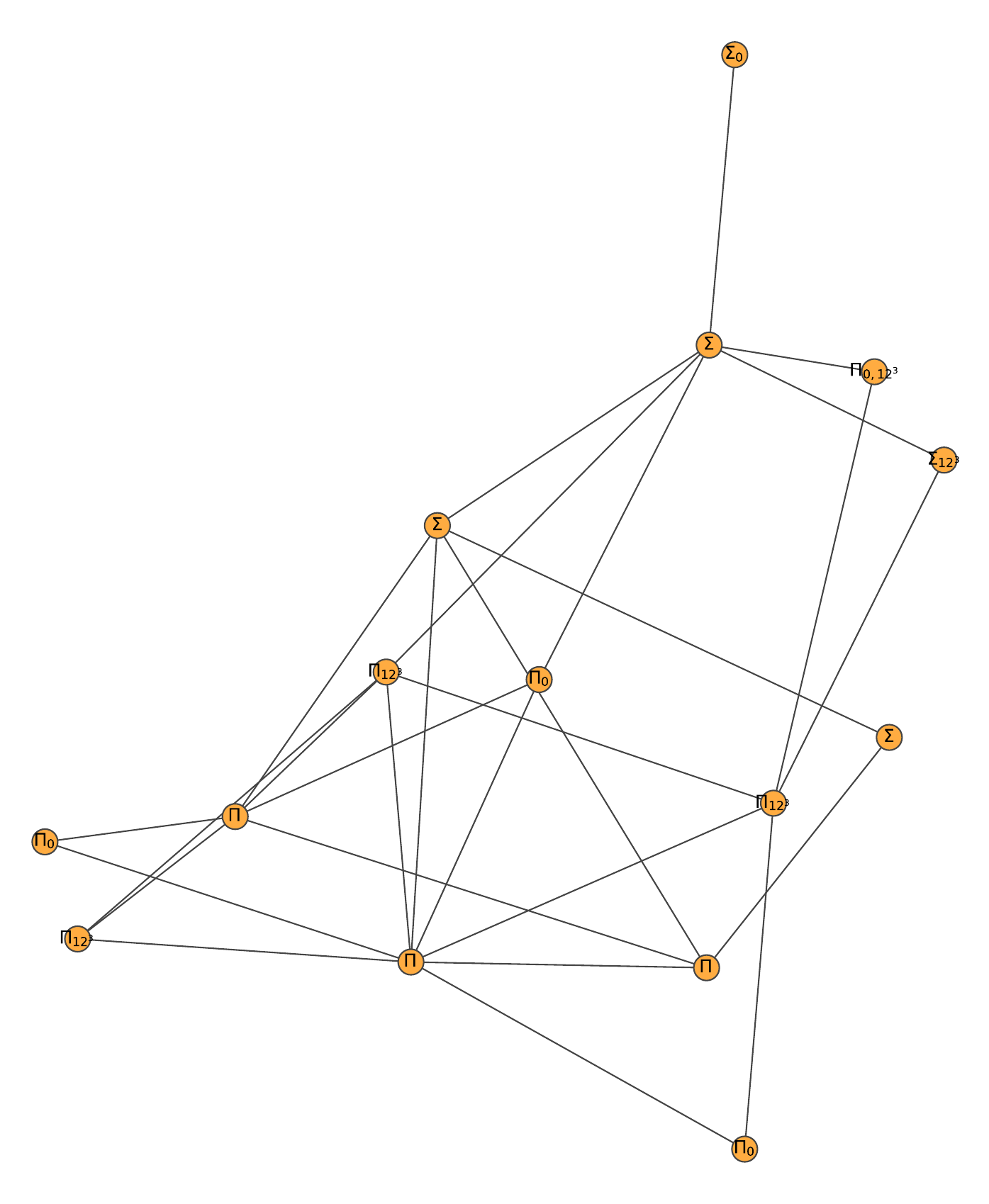} }}%
    \caption{The subgraphs of $\SSIG[2][2][47]$ supported on Jacobians
    (left) and elliptic products (right). Vertex positions are mantained with respect to Figure~\ref{fig:p47}.}%
    \label{fig:p47-subgraphs}%
\end{figure}

In this setting, the vertices which are not of $\TypeA$ in
$\SSIG[2][2][47]^J$ get more isolated, because they all have out-degree less than 15.
On the other hand, the stationary distribution is uniformized slightly in $\SSIG[2][47]^E$,
because the vertices with larger automorphism groups have one, two or three fewer isogenies to Jacobians.
This can be seen in Figure~\ref{fig:p47-subgraphs}. 

These phenomena generalize immediately to $\SSIG[2][\ell][p]$ for all
primes $\ell\neq p$, due to the generality achieved in Theorem~\ref{theorem:isogeny-graphs-distribution}.

%% file: data.tex
\section{
    Experimental diameters and \texorpdfstring{$\lambda_\star$}{λ*} for \texorpdfstring{$\SSIG[2][2]$}{Γ2(2;p)}
}
\label{sec:data}

The following table consists of experimental data computed for the
graphs $G = \SSIG[2][2]$, $J = \SSIG[2][2]^J$ and $E = \SSIG[2][2]^E$.
The computed values are the diameters $d(G)$, $d(J)$ and $d(E)$, and the
(scaled) second-largest eigenvalues of each graph. In particular, the
second eigenvalues of $\SSIG[2][2]$ support Conjecture~\ref{conjecture:eigenvalue-bound}. 
We use the notation $\tilde\lambda_\star = 15\lambda_\star$.

\begin{longtable}{rrrrrrr}
$p$ & $d(G)$ & $d(J)$ & $d(E)$ & $\tilde\lambda_\star(G)$ & $\tilde\lambda_\star(J)$ & $\tilde\lambda_\star(E)$ \\
17  & 3         & 3         & 2         & 10.671       & 9.203        & 3.000        \\
19  & 3         & 3         & 2         & 11.072       & 10.016       & 1.833        \\
23  & 3         & 4         & 2         & 10.241       & 8.993        & 4.102        \\
29  & 4         & 4         & 4         & 10.472       & 9.522        & 6.460        \\
31  & 3         & 4         & 2         & 11.183       & 10.516       & 5.748        \\
37  & 4         & 4         & 2         & 10.797       & 10.025       & 5.372        \\
41  & 5         & 5         & 6         & 11.436       & 10.098       & 7.837        \\
43  & 4         & 4         & 2         & 11.153       & 10.650       & 5.495        \\
47  & 4         & 5         & 4         & 11.131       & 10.526       & 7.580        \\
53  & 5         & 5         & 4         & 11.060       & 10.769       & 6.145        \\
59  & 5         & 5         & 5         & 11.475       & 10.447       & 7.927        \\
61  & 5         & 6         & 3         & 11.451       & 11.037       & 6.978        \\
67  & 5         & 4         & 4         & 11.563       & 11.210       & 7.537        \\
71  & 5         & 5         & 4         & 11.341       & 10.885       & 7.183        \\
73  & 5         & 5         & 4         & 11.577       & 11.129       & 7.575        \\
79  & 5         & 5         & 3         & 11.216       & 10.774       & 6.576        \\
83  & 6         & 6         & 5         & 11.262       & 11.023       & 8.241        \\
89  & 6         & 6         & 6         & 11.307       & 10.681       & 8.418        \\
97  & 5         & 5         & 6         & 11.494       & 11.089       & 7.973        \\
101 & 6         & 6         & 7         & 11.192       & 10.817       & 8.474        \\
103 & 6         & 6         & 5         & 11.217       & 10.980       & 8.644        \\
107 & 6         & 6         & 6         & 11.379       & 11.203       & 7.344        \\
109 & 6         & 6         & 4         & 11.168       & 10.985       & 6.549        \\
113 & 6         & 6         & 6         & 11.386       & 11.156       & 7.593        \\
127 & 6         & 6         & 4         & 11.612       & 11.383       & 7.522        \\
131 & 7         & 6         & 8         & 11.525       & 11.373       & 8.179        \\
137 & 6         & 6         & 6         & 11.648       & 11.440       & 7.193        \\
139 & 6         & 6         & 5         & 11.528       & 11.424       & 7.682        \\
149 & 7         & 7         & 8         & 11.534       & 11.407       & 8.131        \\
151 & 6         & 6         & 4         & 11.387       & 11.285       & 7.338        \\
157 & 6         & 7         & 6         & 11.508       & 11.291       & 8.489        \\
163 & 6         & 6         & 6         & 11.638       & 11.376       & 8.012        \\
167 & 7         & 7         & 6         & 11.494       & 11.359       & 8.116        \\
173 & 7         & 7         & 7         & 11.631       & 11.408       & 8.077        \\
179 & 7         & 7         & 8         & 11.586       & 11.459       & 8.075        \\
181 & 6         & 7         & 6         & 11.347       & 11.267       & 8.270        \\
191 & 7         & 7         & 7         & 11.461       & 11.348       & 8.307        \\
193 & 6         & 6         & 5         & 11.537       & 11.431       & 7.754        \\
197 & 7         & 7         & 8         & 11.295       & 11.207       & 7.789        \\
199 & 7         & 7         & 6         & 11.361       & 11.261       & 8.041        \\
211 & 7         & 7         & 6         & 11.610       & 11.522       & 7.933        \\
223 & 7         & 7         & 7         & 11.484       & 11.339       & 8.334        \\
227 & 7         & 7         & 7         & 11.480       & 11.397       & 8.110        \\
229 & 7         & 7         & 6         & 11.605       & 11.486       & 8.076        \\
233 & 7         & 7         & 6         & 11.523       & 11.420       & 7.672        \\
239 & 8         & 7         & 8         & 11.581       & 11.431       & 8.246        \\
241 & 7         & 7         & 6         & 11.507       & 11.342       & 8.233        \\
251 & 8         & 7         & 8         & 11.568       & 11.371       & 8.585        \\
257 & 8         & 7         & 8         & 11.636       & 11.462       & 8.315        \\
263 & 7         & 7         & 7         & 11.539       & 11.433       & 7.640        \\
269 & 8         & 7         & 8         & 11.448       & 11.337       & 8.405        \\
271 & 7         & 7         & 6         & 11.537       & 11.482       & 8.037        \\
277 & 7         & 8         & 6         & 11.530       & 11.396       & 7.935        \\
281 & 7         & 7         & 8         & 11.479       & 11.366       & 8.297        \\
283 & 7         & 7         & 7         & 11.582       & 11.504       & 8.272        \\
293 & 8         & 8         & 8         & 11.582       & 11.430       & 8.390        \\
307 & 7         & 7         & 7         & 11.614       & 11.535       & 8.244        \\
311 & 8         & 8         & 7         & 11.507       & 11.383       & 8.411        \\
313 & 8         & 7         & 7         & 11.645       & 11.480       & 8.439        \\
317 & 8         & 8         & 7         & 11.543       & 11.495       & 7.922        \\
331 & 7         & 7         & 7         & 11.505       & 11.450       & 8.018        \\
337 & 7         & 7         & 7         & 11.613       & 11.542       & 8.005        \\
347 & 8         & 8         & 8         & 11.520       & 11.457       & 8.185        \\
349 & 8         & 8         & 8         & 11.465       & 11.407       & 8.485        \\
353 & 8         & 8         & 8         & 11.561       & 11.490       & 8.143        \\
359 & 8         & 8         & 8         & 11.556       & 11.500       & 8.311        \\
367 & 8         & 8         & 7         & 11.553       & 11.463       & 8.352        \\
373 & 8         & 8         & 7         & 11.475       & 11.411       & 8.259        \\
379 & 8         & 7         & 7         & 11.474       & 11.408       & 8.202        \\
383 & 8         & 8         & 7         & 11.548       & 11.492       & 8.351        \\
389 & 8         & 8         & 9         & 11.582       & 11.544       & 8.280        \\
397 & 8         & 8         & 7         & 11.593       & 11.523       & 8.368        \\
401 & 8         & 8         & 8         & 11.558       & 11.492       & 8.315        \\
409 & 8         & 8         & 8         & 11.626       & 11.575       & 8.354        \\
419 & 9         & 8         & 10        & 11.555       & 11.472       & 8.552        \\
421 & 8         & 8         & 6         & 11.614       & 11.569       & 8.015        \\
431 & 8         & 8         & 8         & 11.585       & 11.512       & 8.276        \\
433 & 8         & 8         & 9         & 11.615       & 11.532       & 8.516        \\
439 & 8         & 8         & 8         & 11.509       & 11.459       & 8.389        \\
443 & 8         & 8         & 9         & 11.501       & 11.458       & 8.287        \\
449 & 8         & 8         & 8         & 11.546       & 11.499       & 8.178        \\
457 & 8         & 8         & 8         & 11.539       & 11.460       & 8.429        \\
461 & 9         & 8         & 9         & 11.588       & 11.513       & 8.452        \\
463 & 8         & 8         & 9         & 11.514       & 11.458       & 8.394        \\
467 & 8         & 8         & 8         & 11.608       & 11.561       & 8.332        \\
479 & 8         & 8         & 9         & 11.579       & 11.524       & 8.202        \\
487 & 8         & 8         & 8         & 11.546       & 11.512       & 8.320        \\
491 & 8         & 8         & 8         & 11.606       & 11.529       & 8.217        \\
499 & 8         & 8         & 8         & 11.492       & 11.457       & 8.168        \\
503 & 9         & 8         & 8         & 11.606       & 11.529       & 8.209        \\
509 & 9         & 9         & 9         & 11.607       & 11.542       & 8.431        \\
521 & 10        & 8         & 10        & 11.618       & 11.566       & 8.295        \\
523 & 8         & 8         & 8         & 11.596       & 11.545       & 8.338        \\
541 & 8         & 8         & 8         & 11.518       & 11.469       & 8.255        \\
547 & 8         & 8         & 8         & 11.591       & 11.555       & 8.282        \\
557 & 9         & 8         & 10        & 11.528       & 11.490       & 8.277        \\
563 & 9         & 9         & 8         & 11.542       & 11.486       & 8.360        \\
569 & 9         & 8         & 10        & 11.573       & 11.525       & 8.366        \\
571 & 8         & 8         & 8         & 11.605       & 11.560       & 8.262        \\
577 & 8         & 8         & 8         & 11.612       & 11.490       & 8.438        \\
587 & 9         & 9         & 9         & 11.628       & 11.565       & 8.362        \\
593 & 9         & 8         & 10        & 11.642       & 11.565       & 8.446        \\
599 & 9         & 9         & 9         & 11.535       & 11.481       & 8.449        \\
601 & 8         & 8         & 8         & 11.553       & 11.518       & 8.219       
\end{longtable}

%% file: explicit.tex
\section{
    Explicit formul\ae{} for genus-2 computations
}

This appendix collects useful formul\ae{} 
for computing explicit Richelot isogenies,
and identifying the reduced automorphism groups of abelian surfaces.

\subsection{Richelot isogenies}
\label{sec:Richelot}

Let \(\XC: y^2 = F(x)\) be a genus-2 curve,
with \(F\) squarefree of degree \(5\) or \(6\).
The Lagrangian subgroups of \(\Jac{\XC}[2]\)
correspond to factorizations of \(F\) into quadratics
(of which one may be linear, if \(\deg(F) = 5\)):
\[
    \XC: y^2 = F(x) = F_1(x)F_2(x)F_3(x)
    \,,
\]
up to permutation of the \(F_i\) and constant multiples.
We call such factorizations \emph{quadratic splittings}.

Fix one such 
quadratic splitting \(\{F_1,F_2,F_3\}\);
then the corresponding subgroup \(K\subset\Jac{\XC}[2]\) 
is the kernel of a \((2,2)\)-isogeny
\(\phi: \Jac{\XC} \to \Jac{\XC}/K\).
For each \(1 \le i \le 3\),
we write \(F_i(x) = F_{i,2}x^2 + F_{i,1}x + F_{i,0}\).
Now let
\[
    \delta 
    = 
    \delta(F_1,F_2,F_3)
    := 
    \begin{vmatrix}
        F_{1,0} & F_{1,1} & F_{1,2} 
        \\
        F_{2,0} & F_{2,1} & F_{2,2} 
        \\
        F_{3,0} & F_{3,1} & F_{3,2} 
    \end{vmatrix}
    \,.
\]

If \(\delta(F_1,F_2,F_3) \not= 0\),
then 
\(\Jac{\XC}/K\) is isomorphic to a Jacobian \(\Jac{\XC'}\),
which we can compute using Richelot's algorithm
(see~\cite{1988/Bost--Mestre} and~\cite[\S8]{2005/Smith}).
First,
let
\begin{align*}
    G_1(x) & := \delta^{-1}\cdot(F_2'(x)F_3(x) - F_3'(x)F_2(x))
    \,,
    \\
    G_2(x) & := \delta^{-1}\cdot(F_3'(x)F_1(x) - F_1'(x)F_3(x))
    \,,
    \\
    G_3(x) & := \delta^{-1}\cdot(F_1'(x)F_2(x) - F_2'(x)F_1(x))
    \,.
\end{align*}
Now the isogenous Jacobian is \(\Jac{\XC'}\),
where \(\XC'\) is the curve
\[
    \XC': y^2 = G(x) = G_1(x)G_2(x)G_3(x)
\]
and the quadratic splitting \(\{G_1,G_2,G_3\}\)
corresponds to the kernel of the dual isogeny 
\(\dualof{\phi}: \Jac{\XC'}\to\Jac{\XC}\).
The \(F_i\) and \(G_i\) are related by the identity
\[
    F_1(x_1)G_1(x_2)
    +
    F_2(x_1)G_2(x_2)
    +
    F_3(x_1)G_3(x_2)
    +
    (x_1 - x_2)^2
    = 
    0
    \,.
\]
Bruin and Doerksen present a convenient form
for a divisorial correspondence \(\mathcal{R} \subset \XC\times \XC'\)
inducing the isogeny \(\phi\)
(see~\cite[\S4]{2011/Bruin--Doerksen}):
\begin{equation}
    \label{eq:Richelot-defining-equation}
    \mathcal{R}:
    \begin{cases}
        F_1(x_1)G_1(x_2) + F_2(x_1)G_2(x_2) = 0
        \,,
        \\
        F_1(x_1)G_1(x_2)(x_1 - x_2) = y_1y_2
        \,,
        \\
        F_2(x_1)G_2(x_2)(x_1 - x_2) = -y_1y_2
        \,.
    \end{cases}
\end{equation}

If \(\delta(F_1,F_2,F_3) = 0\),
then \(\Jac{\XC}/K\) is isomorphic to 
an elliptic product \(\EC\times\EC'\).
Let \(D(\lambda)\) be the discriminant
of the quadratic polynomial \(F_1 + \lambda F_2\),
and let \(\lambda_1\) and \(\lambda_2\) be the roots of \(D(\lambda)\);
then \(F_1 + \lambda_1 F_2 = U^2\)
and \(F_1 + \lambda_2 F_2 = V^2\)
for some linear polynomials \(U\) and \(V\).
Now \(F_1 = \alpha_1 U^2 + \beta_1 V^2\)
and \(F_2 = \alpha_2 U^2 + \beta_2 V^2\)
for some \(\alpha_1\), \(\beta_1\), \(\alpha_2\), and \(\beta_2\),
and since in this case 
\(F_3\) is a linear combination of \(F_1\) and \(F_2\),
we must have
\(F_3 = \alpha_3 U^2 + \beta_3 V^2\)
for some \(\alpha_3\) and \(\beta_3\).
Now, rewriting the defining equation of \(\XC\) as
\[
    \XC: Y^2 = \prod_{i=1}^3(\alpha_i U^2 + \beta_i V^2)
    \,,
\]
it is clear that 
the elliptic curves
\[
    \EC: Y^2 = \prod_{i=1}^3(\alpha_i X + \beta_i Z)
    \quad
    \text{and}
    \quad
    \EC': Y^2 = \prod_{i=1}^3(\beta_i X + \alpha_i Z)
\]
are the images of double covers
\(\pi: \XC \to \EC\)
and
\(\pi': \XC \to \EC'\)
defined by
\(\pi((X:Y:Z)) = (U:Y:V)\)
and \(\pi'((X:Y:Z)) = (V:Y:U)\),
respectively.
The product of these covers
induces the isogeny \(\phi: \Jac{\XC} \to \EC\times\EC'\).

\subsection{Isogenies from elliptic products}
\label{sec:elliptic-product-isogenies}

Consider a generic pair of elliptic curves over \(\field\),
defined by
\begin{align*}
    \EC: y^2 = (x - s_1)(x - s_2)(x - s_3)
    \shortintertext{and}
    \EC': y^2 = (x - s_1')(x - s_2')(x - s_3')
    \,.
\end{align*}
We have
\( \EC[2] = \{ 0_{\EC}, P_1, P_2, P_3 \} \)
and
\( \EC'[2] = \{ 0_{\EC'}, P_1', P_2', P_3' \} \)
where \(P_i := (s_i,0)\)
and \(P_i' := (s_i',0)\).
For each \(1 \le i \le 3\),
we let 
\[
    \psi_i: \EC \longrightarrow \EC_i := \EC/\subgrp{P_i}
    \quad
    \text{and}
    \quad 
    \psi_i': \EC' \to \EC_i' := \EC'/\subgrp{P_i'}
\]
be the quotient \(2\)-isogenies.
These can be computed using Vélu's formul\ae{}~\cite{1971/Velu}.

The fifteen Lagrangian subgroups of \((\EC\times\EC')[2]\)
fall naturally into two kinds.
Nine of the kernels
correspond to products of \(2\)-isogeny kernels in
\(\EC[2]\).
Namely, 
for each \( 1 \le i, j \le 3 \)
we have a subgroup
\[
    K_{i,j}
    :=
    \subgrp{
        (P_i,0_{\EC'})
        ,
        (0_{\EC},P_i')
    }
    \subset
    (\EC\times\EC')[2]
    \,,
\]
and a quotient isogeny
\[
    \phi_{i,j}: \EC\times\EC' \to (\EC\times\EC')/K_{i,j} \cong \EC_i\times\EC_j'
    \,.
\]
Of course, \(\phi_{i,j} = \psi_i\times\psi_j\);
we can thus compute \(\phi_{i,j}\),
and the codomains \(\EC_i\times\EC_j'\),
using Vélu's formul\ae{} as above.

The other six kernels correspond to \(2\)-Weil anti-isometries
\(\EC[2]\cong\EC'[2]\):
they are
\[
    K_{\pi} 
    := 
    \{ 
        (0_{\EC},0_{\EC'}),
        (P_{1},P_{\pi(1)}'),
        (P_{2},P_{\pi(2)}'),
        (P_{3},P_{\pi(3)}')
    \}
    \quad
    \text{for }
    \pi \in \operatorname{Sym}(\{1,2,3\})
    \,,
\]
with quotient isogenies
\[
    \phi_{\pi}: \EC\times\EC' \to \AV_{\pi} := (\EC\times\EC')/K_{\pi}
    \,.
\]
If the anti-isometry \(P_i \mapsto P_{\pi(i)}'\)
is induced by an isomorphism \(\EC \to \EC'\),
then \(\AV_{\pi}\) is isomorphic to \(\EC\times\EC'\);
otherwise,
it is the Jacobian of a genus-2 curve \(\XC_{\pi}\),
which we can compute using the formul\ae{} 
below (taken from~\cite[Proposition 4]{2000/Howe--Leprevost--Poonen}).

Writing
\(\alpha_i := x(P_{i})\) and \(\beta_i := x(P_{\pi(i)}')\)
for \(1 \le i \le 3\),
let
\begin{align*}
    a_1 & := \frac{(\alpha_3-\alpha_2)^2}{\beta_{3}-\beta_{2}}
           + \frac{(\alpha_2-\alpha_1)^2}{\beta_{2}-\beta_{1}}
           + \frac{(\alpha_1-\alpha_3)^2}{\beta_{1}-\beta_{3}}
    \,,
    \\
    b_1 &:= \frac{(\beta_{3}-\beta_{2})^2}{\alpha_3-\alpha_2}
          + \frac{(\beta_{2}-\beta_{1})^2}{\alpha_2-\alpha_1}
          + \frac{(\beta_{1}-\beta_{3})^2}{\alpha_1-\alpha_3}
    \,,
    \\
    a_2 & := \alpha_1(\beta_{3}-\beta_{2})
            + \alpha_2(\beta_{1}-\beta_{3})
            + \alpha_3(\beta_{2}-\beta_{1}) 
    \,,
    \\
    b_2 & := \beta_{1}(\alpha_3-\alpha_2)
            + \beta_{2}(\alpha_1-\alpha_3)
            + \beta_{3}(\alpha_2-\alpha_1) 
    \,,
    \\
    A & := \Delta'\cdot a_1/a_2
           \text{ where }
           \Delta' := (\beta_{2} - \beta_{3})^2
                      (\beta_{1} - \beta_{3})^2
                      (\beta_{1} - \beta_{2})^2 
    \,,
    \\
    B & := \Delta\cdot b_1/b_2
           \text{ where }
           \Delta := (\alpha_2 - \alpha_3)^2
                     (\alpha_1 - \alpha_3)^2
                     (\alpha_1 - \alpha_2)^2 
    \,,
    \shortintertext{and finally}
    F_1 & := A(\alpha_2-\alpha_1)(\alpha_1-\alpha_3)X^2 
           + B(\beta_{2}-\beta_{1})(\beta_{1}-\beta_{3})Z^2
    \,,
    \\
    F_2 & := A(\alpha_3-\alpha_2)(\alpha_2-\alpha_1)X^2
           + B(\beta_{3}-\beta_{2})(\beta_{2}-\beta_{1})Z^2
    \,,
    \\
    F_3 & := A(\alpha_1-\alpha_3)(\alpha_3-\alpha_2)X^2 
           + B(\beta_{1}-\beta_{3})(\beta_{3}-\beta_{2})Z^2
    \,.
\end{align*}
Now the curve \(\XC_{\pi}\) may be defined by
\[
    \XC_{\pi}: Y^2 = -F_1(X,Z)F_2(X,Z)F_3(X,Z)
    \,.
\]
The dual isogeny \(\dualof{\phi_{\pi}}: \Jac{\XC_{\pi}} \to
\EC\times\EC'\) corresponds 
to the quadratic splitting \(\{F_1,F_2,F_3\}\).

\subsection{Identifying reduced automorphism types of Jacobians}
\label{sec:Clebsch}
We can identify the isomorphism class of a Jacobian \(\Jac{\XC}\)
using the Clebsch invariants \(A\), \(B\), \(C\), \(D\)
of \(\XC\),
which are homogeneous polynomials of degree 2, 4, 6, and 10
in the coefficients of the sextic defining \(\XC\).
These invariants should be seen as coordinates on the weighted projective space
\(\PP(2,4,6,10)\):
that is, 
\[
    (A:B:C:D) = (\lambda^2 A: \lambda^4 B: \lambda^6 C: \lambda^{10} D)
\]
for all nonzero \(\lambda\) in \(\field\).
The Clebsch invariants can be computed using
a series of transvectants involving the sextic 
(see~\cite[\S1]{1991/Mestre}),
but it is more convenient to use
(for example) \texttt{ClebschInvariants} in Magma~\cite{2020/Magma}
or \texttt{clebsch\_invariants} from the
\texttt{sage.schemes.hyperelliptic\_curves.invariants}
library of Sage~\cite{2020/SageMath}.
If \(\XC/\FFbar_p\) is superspecial,
then \((A:B:C:D)\) are in \(\FF_{p^2}\).

To determine \(\RAut(\Jac{\XC})\) for a given genus-2 \(\XC\),
we use necessary and sufficient conditions on the Clebsch invariants
derived by Bolza~\cite[\S11]{1887/Bolza},
given here in Table~\ref{tab:RAut-from-Clebsch}.
These criteria involve some derived invariants:
following Mestre's notation~\cite{1991/Mestre},
let
\begin{align*}
    A_{11} & = 2C + \frac{1}{3}AB
    \,,
    &
    A_{12} & = \frac{2}{3}(B^2 + AC)
    \,,
    &
    A_{23} & = \frac{1}{2}B\cdot A_{12} + \frac{1}{3}C\cdot A_{11}
    \,,
    \\
    A_{22} & = D
    \,,
    &
    A_{31} & = D
    \,,
    &
    A_{33} & = \frac{1}{2}B\cdot A_{22} + \frac{1}{3}C\cdot A_{12}
\end{align*}
(recall again that \(\operatorname{char}\field\) is not \(2\) or \(3\)).
Finally, the \(R\)-invariant is defined by
\[
    R^2 
    = 
    \frac{1}{2}\left|
    \begin{matrix}
        A_{11} & A_{12} & A_{31} \\
        A_{12} & A_{22} & A_{23} \\
        A_{31} & A_{23} & A_{33} \\
    \end{matrix}
    \right|
    \,.
\]

\begin{table}[ht]
    \centering
    \begin{tabular}{|c|c|c|c|}
        \hline
        Type & \(\RAut(\Jac{\XC})\) & Conditions on Clebsch invariants 
        \\
        \hline
        \hline
        \TypeA & \(1\) 
               & \(R \not= 0\), \((A:B:C:D) \not= (0:0:0:1)\)
        \\
        \hline
        \TypeI & \(C_2\)
               & \(R = 0\) and \(A_{11}A_{22} \not= A_{12}\)
        \\
        \hline
        \TypeII & \(C_5\)
                & \((A:B:C:D) = (0:0:0:1)\)
        \\
        \hline
        \multirow{2}{*}{\TypeIII} 
                 & \multirow{2}{*}{\(C_2^2\)}
                 & \(BA_{11} - 2AA_{12} = -6D\),
                 \(D \not= 0\),
        \\
                 &
                 & \(CA_{11} + 2BA_{12} = AD\),
                 \(6C^2 \not= B^3\)
        \\
        \hline
        \multirow{2}{*}{\TypeIV} 
                 & \multirow{2}{*}{\(S_3\)}
                 & \(6C^2 = B^3\),
                   \(3D = 2BA_{11}\),
        \\
                 &
                 & \(2AB \not= 15C\), \(D \not= 0\)
        \\
        \hline
        \TypeV & \(D_{2\times 6}\)
               & \(6B = A^2\),
                 \(D = 0\),
                 \(A_{11} = 0\),
                 \(A \not= 0\)
        \\
        \hline
        \TypeVI & \(S_4\)
                & \((A:B:C:D) = (1:0:0:0)\)
        \\
        \hline
    \end{tabular}
    \caption{The classification of reduced automorphism groups of
        Jacobian surfaces, with necessary and sufficient conditions on 
        the Clebsch invariants for each type.
    }
    \label{tab:RAut-from-Clebsch}
\end{table}